\documentclass[useAMS,usenatbib,usegraphicx]{mn2e}

\usepackage{hyperref}
\usepackage{color}
\usepackage{amsmath}
\usepackage{aas_macros}
\usepackage{units}

\begin{document}

\title[Calculating the transfer function of PCA]{Calculating the transfer function of noise removal by principal component analysis and application to AzTEC deep-field observations}
\author[T.P. Downes et al.]{T.P. Downes$^1$, D. Welch$^2$, K.S. Scott$^3$, J. Austermann$^4$, G.W. Wilson$^2$, M.S. Yun$^2$\\
$^1$California Institute of Technology, Pasadena, CA 91125 USA\\
$^2$Department of Astronomy, University of Massachusetts, Amherst, MA 01003, USA\\
$^3$North American ALMA Science Center, National Radio Astronomy Observatory, Charlottesville, VA 22903\\
$^4$Center for Astrophysics and Space Astronomy, University of Colorado, Boulder, CO 80309, USA}
\date{Accepted 2012 MM DD. Received 2011 March DD.}
\maketitle

\begin{abstract}
Instruments using arrays of many bolometers have become increasingly common in the past decade. The maps produced by
such instruments typically include the filtering effects of the instrument as well as those from subsequent steps performed in the 
reduction of the data. Therefore interpretation of the maps is dependent upon accurately calculating the transfer function of the 
chosen reduction technique on the signal of interest. Many of these instruments use non-linear and iterative techniques to reduce 
their data because such methods can offer improved signal-to-noise over those that are purely linear, particularly for signals at 
scales comparable to that subtended by the array. We discuss a general approach for measuring the transfer function of principal 
component analysis (PCA) on point sources that are small compared to the spatial extent seen by any single bolometer within the 
array. The results are applied to previously released AzTEC catalogues of the COSMOS, Lockman Hole, Subaru XMM-Newton 
Deep Field, GOODS-North and GOODS-South fields. Source flux density and noise estimates increase by roughly  $+10$ per 
cent for fields observed while AzTEC was installed at the Atacama Submillimeter Telescope Experiment and +15-25 
per cent while AzTEC was installed at the James Clerk Maxwell Telescope. Detection significance is, on average, unaffected by 
the revised technique. The revised photometry technique will be used in subsequent AzTEC releases.
\end{abstract}


\section{Introduction}
The development of instruments with arrays of 100 to several 1000 bolometers to detect submillimeter and millimeter 
radiation has become commonplace in the past decade. For ground-based instruments, the predominant signal in the 
recorded data is emission from the atmosphere, particularly that from precipitable water vapor. Other undesirable features
in the recorded data include noise -- features in the data not traceable to incoming photons -- that are introduced via a
number of mechanisms. We can hope to remove sources of noise which are correlated from one detector to at least one
other, though many are common to the whole array or portions within. The removal of these undesirable, correlated features
from the data is a major hurdle in reducing the recorded data into cleaned data that are hopefully dominated by astrophysical
signal  and unremovable random noise. In a typical AzTEC \citep{WilsonAztecInstrument} observation, $\sim$90 per cent of
the atmospheric emission is described by the average signal across the array. Thus the primary problem is to ``clean'' the
recorded data of the remaining atmospheric signal at higher moments and other sources of correlated noise without removing
signal to a degree that degrades signal-to-noise.

Techniques which target and remove specific modes from the recorded data are commonly used because they are
typically based upon models which incorporate physical phenomena and an understanding of the instrument.
In particular, linear techniques may be preferred because they have the distributive and scalar multiplicative 
properties of linear operators. That is, if one knows how the signal of interest will manifest in the recorded data,
then one can estimate the filtering effect of the cleaning technique -- the ``transfer function'' -- on simulated data which
contain only such a signal, without being compelled to use real or simulated atmospheric signal and other sources of
noise. The final map can be appropriately normalized by this transfer function to produce a result in astrophysical units.
It is found, however, that purely linear techniques often provide unsatisfactory signal-to-noise performance in that they
remove insufficient noise or too much signal, particularly for signals that subtend a significant fraction of the array.

Thus non-linear, sometimes iterative, techniques have been developed to improve detection significance
\citep{Enoch2006, KovacsCRUSH2008, SayersMK2010}. The need for this development can be understood from some 
simple properties of real-world instruments without resorting to measuring or modeling the properties of the atmospheric 
emission (\textit{e.g.}, \cite{LayHalverson2000, SayersMK2010}) or any other undesirable feature in the recorded data.
An actual instrument employs detectors whose response to sky signal (both atmospheric and astrophysical in origin) may
be a varying function of time owing to, \textit{e.g.}, the nature of the detection mechanism, variation in the subsequent
electronic amplification or changes in the optical properties of the instrument. Though instruments are calibrated at regular
intervals, variations on time scales much shorter than the interval cannot be accounted for in the calibration. The relative
gain between detectors is important because modeling the largest undesirable feature, the atmosphere, requires converting
the recorded data to values proportional to physical units. A small fluctuation, or calibration imprecision, in the relative gain
between detectors can have significant impact because it is multiplied by the large correlated atmospheric signal. Allowing
the relative gains used in atmospheric removal to converge to a set of values independent of the  calibration values, as in
\cite{SayersMK2010}, is an example of a non-linear technique because the data themselves are used to measure the
relative gain; \textit{i.e.}, the cleaned data are a function of the recorded data multiplied by the relative gain, which is no
longer independent of the data. Similarly, principal component analysis (PCA) allows the relative gain between detectors
to be determined by the covariance matrix calculated from the recorded data and is thereby non-linear.

To interpret a map produced by a non-linear analysis technique, we still require a transfer function for the signal of 
interest. Though non-linear techniques will not, in general, have the distributive and multiplicative properties of
linear operators, the interpretation of the map depends only on the cleaned signal of interest being linearly
proportional to the input recorded data. Though it is non-linear, the technique described in \cite{KovacsCRUSH2008}
retains the transfer function estimation advantages of linear techniques because the data is modeled explicitly as the
summation of specified noise modes and astrophysical signal. The PCA technique, described further in Sec. \ref{sec:pca},
``adaptively'' uses the recorded data to identify the modes to be removed. Thus, calculation of a PCA transfer function must
ultimately rely on the recorded data themselves.

We describe herein applications of this approach to PCA on data from the AzTEC instrument and make comparisons to
approximations to the full non-linear problem. The resulting photometry is applied to the AzTEC data and revised versions of
previously released catalogues are presented. No changes to the correlated noise removal technique itself are made.

\section{Principal component analysis}
\label{sec:pca}
Principal component analysis is a popular technique for identifying the moments that describe the variance
in data without relying on having measured those moments in their natural coordinate frame. As applied to bolometric
arrays, the recorded data from $N$ bolometers with $N_\text{samples}$ each are decomposed into orthonormal eigenfunctions
by the standard eigen-decomposition technique (\textit{e.g.}, \cite[chap. 7]{linalg}). The eigenfunctions can be rank-ordered in 
eigenvalue and thus also by their contribution to the variance in the recorded timestream. The largest eigenfunctions are
then supposed to have their origins in atmospheric signal as well as other strong correlations in the instrument. Since these
modes are determined by the data themselves, the process as a whole is non-linear, even though eigen-decomposition and
eigenfunction removal are individually linear. 

The exact choice of the number of eigenfunctions to remove from the recorded data is somewhat arbitrary. It is empirically observed 
that a logarithmic distribution of the eigenvalues will contain a large cluster of low eigenvalues\footnotemark~along with a 
number of widely distributed larger eigenvalues. In the AzTEC pipeline, the width of the low-eigenvalue cluster is used to 
calculate the number of eigenfunctions to remove. Though other cuts could be made, this choice allows a simple parameter, a 
multiplier on the eigenvalue distribution width, to control the cleaning process. Eigenfunctions are removed from the data until no 
further modes exist outside a region defined by the multiplier times the distribution width. It is observed that the
number of modes removed is unaffected upon addition of simulated sources of typical flux densities (a few to 10 mJy at 1.1 mm) 
to the recorded data. The particular value of 2.5 for this multiplier has been empirically found to roughly maximize signal-to-noise for point sources. Typically 5-15 modes are removed from the data. The details of this technique are described in
\cite{ScottCOSMOS2008}.
\footnotetext{Simulated data that contain only random noise have such a feature, suggesting its origin.}

The advantage of the PCA technique is that the largest correlations are adaptively identified and removed. This removes
large correlated features that may be easily described by physically motivated models as well as features that do not lend
themselves to modeling. An example of the latter might be electromagnetic interference that couples to detectors
with a strength that varies with time or is found only in a subset of the data. By automatically removing these features, the observer's
time can be dedicated to interpretation of the interesting signals. However, the transfer function of PCA on signals is
dependent on what modes are adaptively identified and removed. The transfer function
estimation technique described in \cite{ScottCOSMOS2008} (and used in subsequent AzTEC publications) is 
a linear approximation to the PCA cleaning operator because it assumes that the operator -- which identifies high power 
modes that are correlated between detectors -- is unaffected by the presence of a simulated faint source. We might expect this 
to be true because point sources subtend an angle that is small compared to the bolometer spacing and also because the 
typical signal they contribute is small compared to that from the atmosphere, but it is not empirically observed to be true.

In fact, the eigenfunction spectrum at large eigenvalue is systematically affected by the addition of simulated sources of
typical flux densities to the recorded data. Comparing the eigenvectors\footnotemark~calculated from the recorded and source-added 
data, we find that the cleaning operator components vary at the several percent level (with roughly equal fluctuations 
upwards and downwards) for the largest eigenvalue eigenfunction. Because the largest eigenfunction is essentially the average 
atmospheric signal across the array \citep{SayersThesis}, a several percent effect in the operator can be significant compared to 
the source flux we intend to measure.
\footnotetext{The eigenvectors are an $N \times N$ matrix that transform the recorded data into the orthonormal basis of 
eigenfunctions. Any changes in the eigenvector components corresponding to large eigenvalues are reflected in the PCA 
cleaning operator that removes eigenfunctions from the data.}

This observation calls into question the accuracy of a linear approximation to the PCA cleaning operator. A full, non-linear 
simulation of the cleaning operator is therefore necessary. As will be shown, the linear approximation results in a systematic 
overestimation of the transfer function and an underestimate of the flux and noise present in the optimally filtered map.

\section{Simulation of the PCA Transfer Function}
If we clean and map data with simulated sources and difference them from unfiltered maps produced from the recorded 
data, we can see how point sources are affected by PCA cleaning. For the purposes of this analysis, we have chosen
to apply this technique to several previously published AzTEC deep-field observations (described in greater detail in Sec.
\ref{sec:catalogues}) of size varying from $\sim$0.1 to 0.4 square degrees.

Prior to performing the simulation, we produce an initial filtered map using the linear prescription in \cite{ScottCOSMOS2008}. This
map can be used to estimate the final noise 
level and to calculate a region of the map that will be used in subsequent analysis. Typically this region is defined by including 
pixels whose noise-weighted time coverage is 50-70 per cent of the maximum coverage in the map. Simulated source locations 
are chosen to be more than 60\arcsec~away from sources detected with significance greater than 3.5 in the selected region of the 
initial map. Likewise, all simulated sources are chosen to have a flux density equal to 10 times the average noise level of the 
selected region in the initial map. For each field, we insert 3 simulated sources per 0.05 square degrees with a maximum of 8. 
These 3 choices ensure that the transfer function is measured on simulated sources that are comparable to typically observed 
sources but are not affected by the true bright sources and do not themselves strongly affect the data.

The noise realisations in the recorded and source-added maps are similar but not precisely the same because the distributive
property does not hold for non-linear operators. Thus, simple differencing of the maps is insufficient to produce
a proper transfer function because it will include residual noise on pixel scales that is not a reflection of the actual effect of 
cleaning on a point source signal. This residual is typically small compared to the noise level in the map; however, its use in
an optimal filter would wrongly couple noise into our estimate of source flux and detection significance. We mitigate this effect 
through 3 additional steps: (1) stacking the difference map at the centre of the simulated sources and normalising by the
known inserted flux, (2) rotationally averaging the stacked signal, and (3) tapering the stacked signal at a distance 4 times the 
FWHM from the beam centre. These steps must ultimately be justified \textit{a posteriori} -- do they produce a transfer function that 
works? However, rotational averaging can be justified \textit{a priori} through an understanding of the AzTEC observing strategy.
AzTEC maps are produced by co-adding many individual maps (typically 60 or greater) taken at many elevations. Each individual
scan, whether raster or Lissajous, is performed in azimuth and elevation while tracking a fixed centre point. The software tracks
position angle of the beam and can detect when pixels are weakly cross-linked. Given this observing strategy, we expect that
the point source transfer function should exhibit significant cylindrical symmetry. Likewise, tapering the signal at the edges is
justified because any measured difference is unlikely to be physical in origin. 

\begin{figure*}
\includegraphics[height=0.95\textwidth, angle=-90]{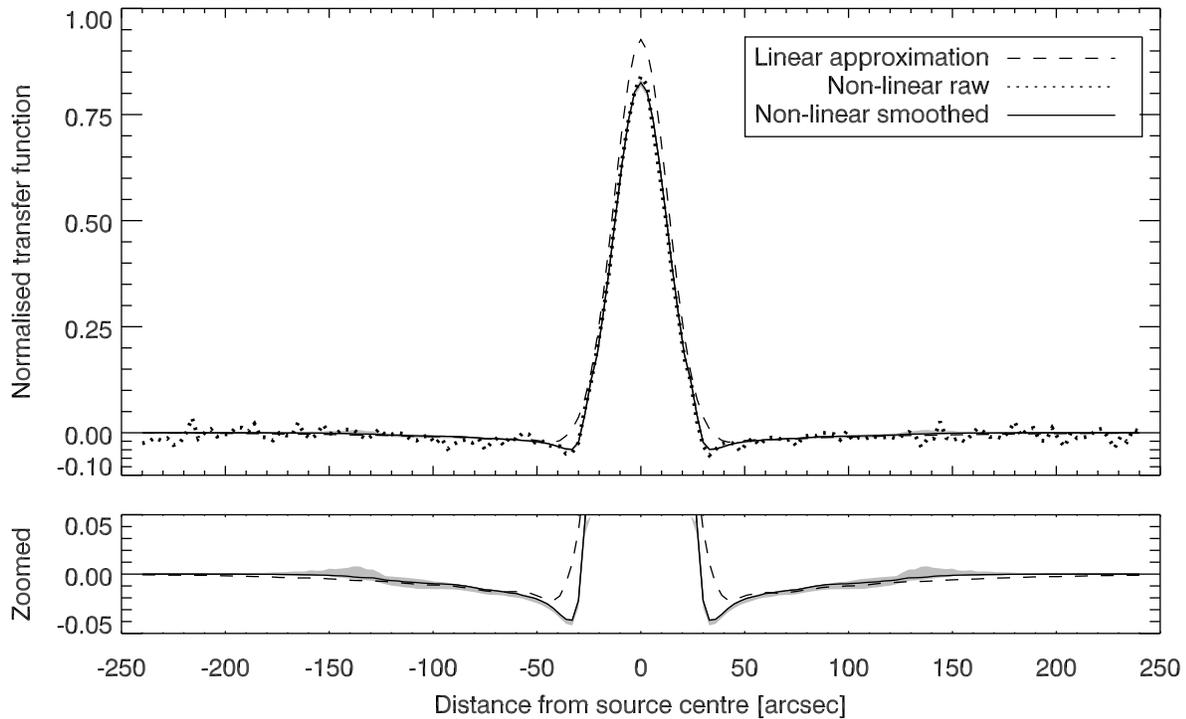}
\caption{Comparison of the various techniques to estimate the PCA cleaning point source transfer function for a set of 43
observations of a single ASTE science field. See text for full discussion of interpretation. The grey shaded region indicates the 
envelope of 20 calculations of the non-linear smoothed kernel using different locations for the simulated sources.}
\label{fig:stacking}
\end{figure*}

\begin{figure*}
\includegraphics[width=0.95\textwidth]{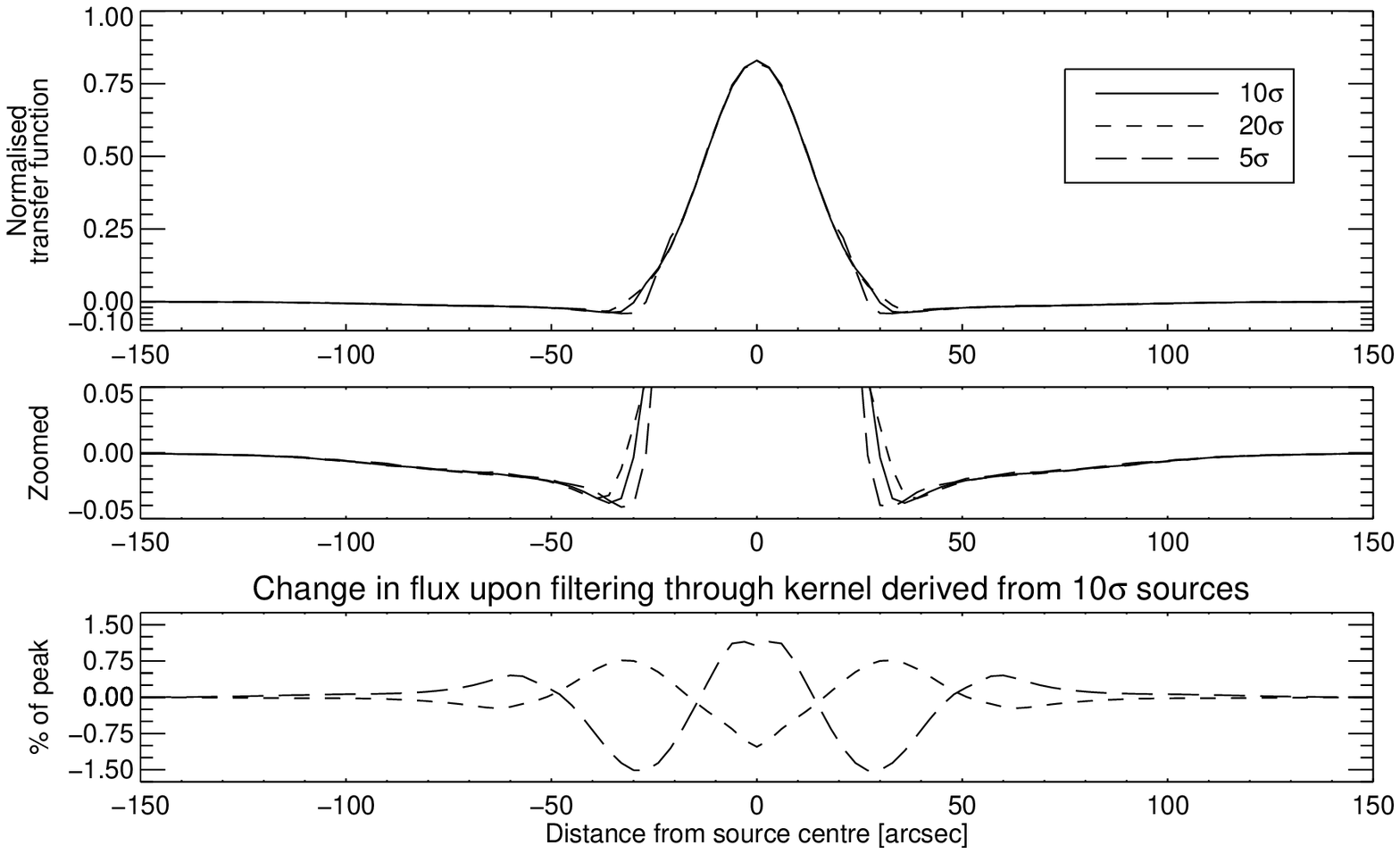}
\caption{(a) Comparison of transfer functions calculated using 3 simulated sources of varying brightness at fixed locations. (b)
A zoomed portion of (a) showing that the primary impact of non-linearity is to make the negative sidelobes shallower as source
brightness increases. (c) The change in measured flux (relative to peak) for $20\sigma$ and $5\sigma$ sources when they are
optimally filtered using the the standard transfer function derived from $10\sigma$ sources. The peak flux and all other pixels
are shifted by $<1.5\%$ of the peak flux and therefore the transfer function does not introduce significant systematic error for
sources of typical detection significance.}
\label{fig:intensity}
\end{figure*}

In Fig. \ref{fig:stacking}, we show a cut in elevation through the transfer function estimates for the previously selected field 
resulting from the linear approximation, differencing/stacking, and differencing/stacking with the extra steps noted above. The 
transfer function for each field will be slightly different owing to differing observing conditions and the non-linear nature of PCA 
cleaning. This transfer function is representative of typical values seen for observations from the Atacama Submillimeter 
Telescope Experiment (ASTE). It is seen that the linear approximation overestimates the peak signal and underestimates the 
negative sidelobes that result from the effective high-pass filter of the cleaning operator. The lower peak value and larger 
sidelobes can be understood as accounting for the effect of the source itself on the atmospheric model; the cleaning operator 
mistakenly includes some source flux in its atmospheric removal thus reducing the peak signal and increasing the sidelobes 
(which result in part from the source's contribution to the array average signal). Differencing and stacking simulated sources 
results in a more accurate transfer function estimation, albeit with imperfect differencing of noise. This is effectively resolved by 
rotationally averaging and tapering the map far from the source centre.  The process was repeated many times using sources
at varying locations and produces a stable result, as indicated by the small scatter in values around the particular realisation
presented. Furthermore, this analysis was reproduced using 3 simulated sources of varying brightness at fixed locations
(Fig \ref{fig:intensity}). It is seen that the primary impact of non-linearity on the transfer function is to make the negative
sidelobes shallower as source brightness increases. The impact on measured flux is negligible for sources of typical detection
significance.

\begin{figure}
\includegraphics[width=0.95\columnwidth]{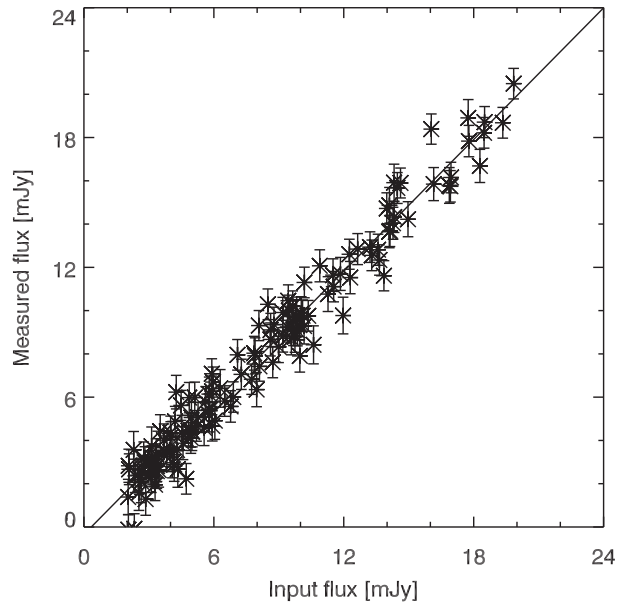}
\caption{A comparison of the observed flux in optimally filtered maps at the locations where simulated sources of known flux
have been inserted. The transfer function is consistent with unity and has a small, negative offset which can be explained by observing
that the mean value of the chosen pixel locations was -0.24 mJy in the recorded map. The best-fit line,
$y = -0.3 \pm 0.2 + (1.015\pm0.018)x$, is shown in red.}
\label{fig:inout}
\end{figure}

The ultimate test of the revised transfer function is whether it succeeds in producing the correct flux when analyzing
data with simulated sources reduced blindly. 23 simulated maps were produced in each of which were inserted 4 simulated 
sources at varying locations far from resolved sources with fluxes ranging from 3 to 20mJy. This spans a detection significance 
range of $\sim$$4-30\sigma$. The sources were placed at the centre of 3\arcsec~pixels in the portion of the map with sufficient 
and uniform coverage to be used for selecting true astrophysical sources. This simulation also tests for any impact that a
moderate increase in source density may have upon the transfer function as 7 simulated sources will ultimately be inserted into
the map (4 ``test'' sources whose flux we intend to measure and 3 sources whose sole purpose is to measure the transfer function).
The maps were then optimally filtered using the revised transfer function estimate and the detected flux and estimated noise at the
known source location was compared to the known input flux (Fig. \ref{fig:inout}). The input and observed fluxes are found to be
consistent; there is a small negative offset that is consistent with the mean value (-0.24 mJy) of the pixels at the chosen input
locations. Because the map has an overall mean of zero and we have chosen locations that are far from bright, positive sources
of flux, it is reasonable to find a small, negative offset. The absence of systematic effects from source location (Fig. \ref{fig:stacking})
or source flux density (Fig. \ref{fig:inout}) may be an indication that, although the transfer function must be varying as a function of
time (the number of eigenfunctions removed from each chunk of data is not constant), it varies more slowly than the time taken to
cover the useful coverage region of the maps. Thus the variations are captured equally well by any simulated source within this
region.

\begin{figure*}
\includegraphics[height=0.95\textwidth, angle=-90]{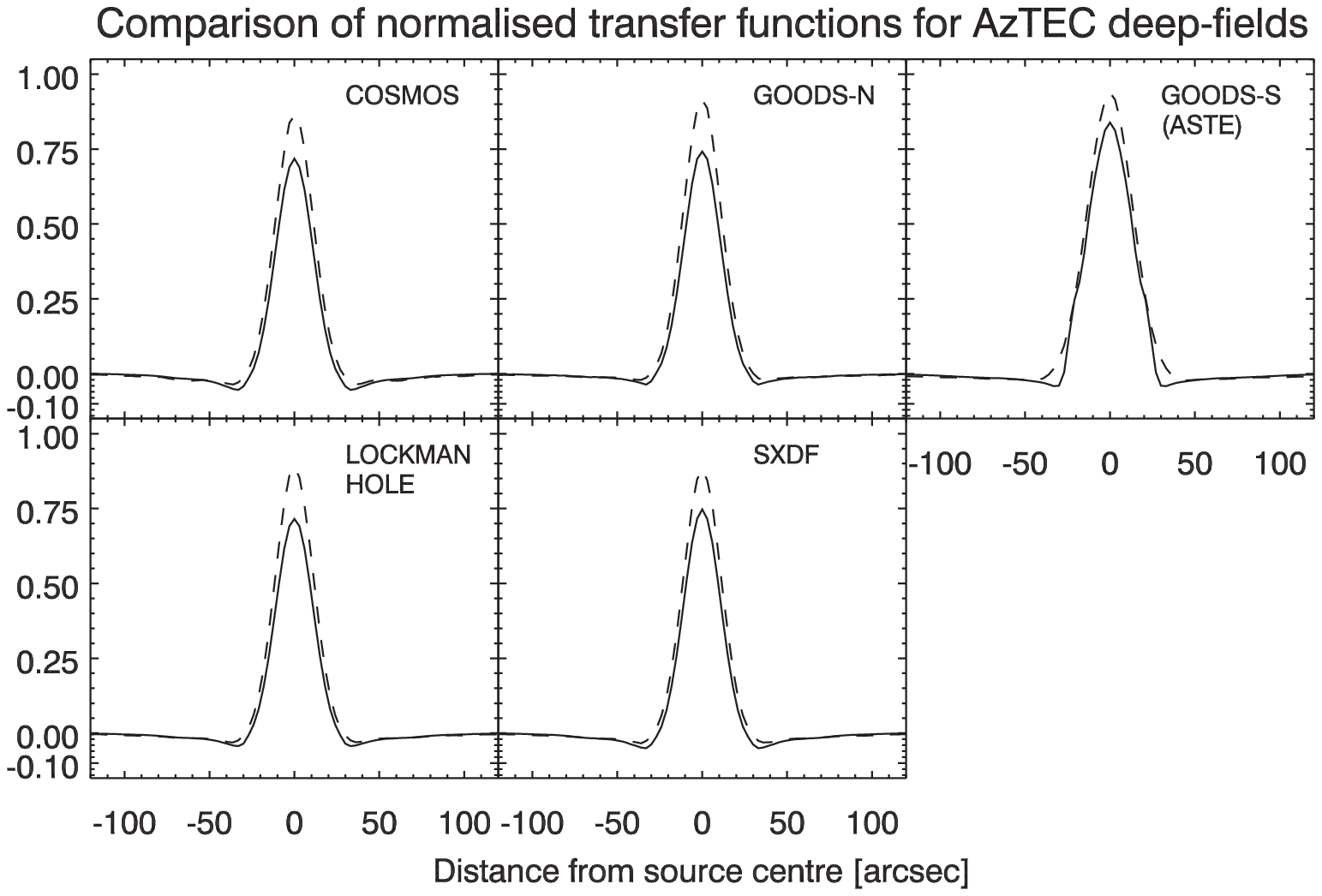}
\caption{Comparison of the linear approximation (dashed) and simulated source (solid) techniques to estimate the PCA cleaning
point source transfer function for previously published AzTEC catalogues. It is observed that the impact is greater for the 4 JCMT
fields than for the ASTE field shown as well as in internal analysis for ASTE fields not yet published. This may be due to the greater
atmospheric fluctuations at the JCMT site.}
\label{fig:allfields}
\end{figure*}

\section{Revised catalogues}
\label{sec:catalogues}
When this technique is applied to other fields (Fig. \ref{fig:allfields}), we observe, for AzTEC data taken while installed at the
ASTE, that the revised transfer function corrects both signal and noise by $\sim$+10 per cent, in each field. Similarly, we find a
correction of +15-25 per cent in the fields observed while AzTEC was installed at the James Clerk Maxwell Telescope (JCMT).
This is consistent with the notion that some of the point source signal is present in the largest eigenfunctions removed by PCA.
The larger impact for the JCMT data may be taken as a sign that the non-linear nature of PCA is more greatly affected by the
worse observing conditions at Mauna Kea as compared to the Atacama Desert in Chile.  We wish to emphasize that,
for each observation, simulating the signal of interest and directly observing the impact of PCA or another non-linear technique
is a surer approach than building expectations based on prior results.

Several previous publications have released point source catalogues from the AzTEC instrument while it was installed at the  
JCMT and the ASTE. These catalogues are reproduced below using the revised photometry, along with deboosted fluxes
calculated from a forthcoming number counts analysis to be presented in Scott, \textit{et al}., submitted\footnote{The best-fit parameters for the observed blank-field number counts were found to be $N_\text{3mJy} = 231 \unit{mJy}^{-1} \cdot \unit{deg}^{-2}$ and $S' = 1.84 \unit{mJy}$ while fixing $\alpha \equiv -2$ and following Eqns. 2 and 3 in \cite{AustermannSHADES2010}.}.
The AzTEC deboosting algorithm \citep{AustermannCOSMOS2009} accounts for the fact that, for a source population
that declines steeply with flux, any given source is more likely to be a relatively plentiful dim source ``boosted'' upward by
noise than a rare bright source on
top of a negative noise fluctuation. The algorithm makes the assumption that the flux in a given pixel is emitted from a single
source. This assumption may be invalid owing to source confusion from the finite AzTEC beam, however it has been shown
to yield results consistent with a parametric frequentist ``$P(d)$'' approach \citep{PereraGOODSN2008} that attempts to account
for confusion through comparison of the recorded map to simulations of noise and source populations convolved with the beam. The
number counts analysis incorporates the revised photometry and is based upon measurements in all blank fields, including
those below, surveyed by the AzTEC instrument. Because the effect of deboosting is a slowly changing function of the sky
model prior assumed, the deboosted values given below are reasonable even if the inclusion of other fields introduces a
potential bias to the sky model. The deboosted fluxes also include correction for a bias introduced by searching for a signal
peak in the presence of noise. The number counts analyses previously presented will be affected by the change in photometry
in 2 ways. First, the estimated counts at a given brightness value are now appropriate for a slightly increased source brightness.
Second, an integral step in computing number counts is dividing by the estimated completeness\footnotemark~for a given flux
in the map. Because the estimated error is higher under the revised transfer function, any given flux value will be less complete;
this effect will be more pronounced at moderate signal-to-noise ($\sim$3-5) where the completeness is rapidly dropping from unity
toward zero.
\footnotetext{The likelihood of detecting a source of a particular brightness given the sources of noise present in the map.}

The details of the analysis for each field do not differ appreciably from that previously performed, except for the change in 
photometry. We provide references to these analyses for the interested reader. We present the catalogues using the
significance and spatial cuts applied by each catalogue's respective author. In each case the expected
false detection rate is estimated using the defined cuts and did not change greatly from that using the transfer function
derived from the linear approximation. Source names and numeric identifiers
for sources which were previously detected are retained; however a common format has been chosen: ``AzTEC/field\#'', where
`\#' indicates order of discovery rather than strictly being based upon detection significance in these revised catalogues. Thus newly 
discovered sources with higher significance than a previously discovered source will appear at the end of these catalogs and a slight 
shuffling (in significance) of previously discovered sources will occur for the reasons discussed above.

\subsection{COSMOS / JCMT}
The COSMOS survey \citep{ScottCOSMOS2008, AustermannCOSMOS2009} was undertaken by the AzTEC instrument in 2005 
while it was installed at the JCMT. A revised catalogue is shown in Table \ref{tbl:cosmos}. A spatial cut (0.15 deg$^2$) was applied
by taking only pixels within the map whose weighting (a combination of noise in the data and amount of time spent observing that
pixel) are greater than 75 per cent of the map's characteristic value (roughly the maximum). A significance cut is applied by taking only
sources whose signal to noise ratio are greater than 3.5. We have conservatively over-estimated the false detection rate to be 25 per cent
($\sim$12 sources) using the number of sources ``detected'' in pure noise realisations of the field. This choice is conservative because
faint sources contribute to flux in nearly every pixel in the map and therefore the likelihood of finding a source at any pixel is greater than
it otherwise would be \citep{PereraGOODSN2008,ScottGOODSS2010}. The difference between this technique and one that attempts
to account for this effect through simulations of source populations can be a factor of 2 or greater.

\begin{table*}
\caption{The AzTEC point source catalogue for the COSMOS field as observed from the JCMT.}
\begin{tabular}{|l|l|c|c|c|c|c|c|c|}
\hline \hline
Source ID & Nickname & S/N & $S_{1.1\text{mm}}$ & $S^\text{corrected}_{1.1\text{mm}}$ & $P(<0)$ & Flux & Noise & $\theta$\\
& & & [mJy] & [mJy] & & increase & increase & \\
\hline
AzTEC\_J095942.68+022936.1 & AzTEC/COSMOS.1 &  8.1 & $12.4 \pm 1.5$ & $10.8 \pm ^{+ 1.5}_{- 1.6}$ &  0.00 & 16.0\% & 18.9\% &  0.6\arcsec \\
AzTEC\_J100008.03+022612.0 & AzTEC/COSMOS.2 &  7.3 & $11.3 \pm 1.5$ & $ 9.7 \pm ^{+ 1.4}_{- 1.7}$ &  0.00 & 16.8\% & 18.8\% &  0.6\arcsec \\
AzTEC\_J100018.26+024830.1 & AzTEC/COSMOS.3 &  6.5 & $10.7 \pm 1.6$ & $ 8.6 \pm ^{+ 1.7}_{- 1.7}$ &  0.00 & 20.9\% & 18.8\% &  1.0\arcsec \\
AzTEC\_J100006.40+023839.9 & AzTEC/COSMOS.4 &  6.2 & $ 9.0 \pm 1.5$ & $ 7.3 \pm ^{+ 1.5}_{- 1.5}$ &  0.00 & 17.1\% & 18.8\% &  0.3\arcsec \\
AzTEC\_J100019.73+023205.8 & AzTEC/COSMOS.5 &  6.0 & $ 9.1 \pm 1.5$ & $ 7.3 \pm ^{+ 1.6}_{- 1.5}$ &  0.00 & 15.4\% & 18.8\% &  0.6\arcsec \\
AzTEC\_J100020.72+023518.3 & AzTEC/COSMOS.6 &  5.9 & $ 8.8 \pm 1.5$ & $ 6.9 \pm ^{+ 1.6}_{- 1.4}$ &  0.00 & 18.8\% & 18.8\% &  0.7\arcsec \\
AzTEC\_J095959.33+023445.8 & AzTEC/COSMOS.7 &  5.4 & $ 8.0 \pm 1.5$ & $ 6.0 \pm ^{+ 1.7}_{- 1.4}$ &  0.00 & 12.8\% & 18.8\% &  0.5\arcsec \\
AzTEC\_J095957.22+022729.3 & AzTEC/COSMOS.8 &  5.5 & $ 8.4 \pm 1.5$ & $ 6.5 \pm ^{+ 1.6}_{- 1.6}$ &  0.00 & 16.7\% & 18.8\% &  1.2\arcsec \\
AzTEC\_J095931.82+023040.1 & AzTEC/COSMOS.9 &  5.0 & $ 7.5 \pm 1.5$ & $ 5.5 \pm ^{+ 1.6}_{- 1.6}$ &  0.00 & 12.3\% & 18.7\% &  0.6\arcsec \\
AzTEC\_J095930.76+024034.2 & AzTEC/COSMOS.10 &  5.1 & $ 7.3 \pm 1.4$ & $ 5.5 \pm ^{+ 1.4}_{- 1.6}$ &  0.00 & 17.8\% & 18.7\% &  0.7\arcsec \\
AzTEC\_J100008.79+024008.0 & AzTEC/COSMOS.11 &  5.1 & $ 7.3 \pm 1.4$ & $ 5.5 \pm ^{+ 1.5}_{- 1.5}$ &  0.00 & 18.9\% & 18.8\% &  0.5\arcsec \\
AzTEC\_J100035.37+024352.3 & AzTEC/COSMOS.12 &  4.9 & $ 7.5 \pm 1.5$ & $ 5.5 \pm ^{+ 1.5}_{- 1.7}$ &  0.00 & 22.2\% & 18.7\% &  1.0\arcsec \\
AzTEC\_J095937.05+023315.4 & AzTEC/COSMOS.13 &  4.6 & $ 6.9 \pm 1.5$ & $ 4.9 \pm ^{+ 1.5}_{- 1.7}$ &  0.00 & 15.2\% & 18.8\% &  0.9\arcsec \\
AzTEC\_J100010.00+023021.2 & AzTEC/COSMOS.14 &  4.8 & $ 7.3 \pm 1.5$ & $ 5.2 \pm ^{+ 1.6}_{- 1.6}$ &  0.00 & 21.3\% & 18.8\% &  1.7\arcsec \\
AzTEC\_J100013.22+023428.1 & AzTEC/COSMOS.15 &  4.4 & $ 6.5 \pm 1.5$ & $ 4.5 \pm ^{+ 1.5}_{- 1.7}$ &  0.00 & 13.0\% & 18.7\% &  0.5\arcsec \\
AzTEC\_J095950.29+024416.2 & AzTEC/COSMOS.16 &  4.5 & $ 6.3 \pm 1.4$ & $ 4.5 \pm ^{+ 1.4}_{- 1.6}$ &  0.00 & 17.5\% & 18.7\% &  0.5\arcsec \\
AzTEC\_J095939.29+023408.2 & AzTEC/COSMOS.17 &  4.4 & $ 6.5 \pm 1.5$ & $ 4.5 \pm ^{+ 1.5}_{- 1.7}$ &  0.01 & 18.7\% & 18.8\% &  0.0\arcsec \\
AzTEC\_J095943.05+023540.2 & AzTEC/COSMOS.18 &  4.3 & $ 6.3 \pm 1.5$ & $ 4.3 \pm ^{+ 1.6}_{- 1.7}$ &  0.01 & 18.2\% & 18.8\% &  0.5\arcsec \\
AzTEC\_J100028.93+023200.3 & AzTEC/COSMOS.19 &  4.4 & $ 6.6 \pm 1.5$ & $ 4.5 \pm ^{+ 1.6}_{- 1.7}$ &  0.00 & 22.4\% & 18.8\% &  0.2\arcsec \\
AzTEC\_J100020.16+024117.2 & AzTEC/COSMOS.20 &  4.0 & $ 5.8 \pm 1.4$ & $ 3.7 \pm ^{+ 1.6}_{- 1.6}$ &  0.01 & 12.0\% & 18.6\% &  2.0\arcsec \\
AzTEC\_J100002.73+024645.0 & AzTEC/COSMOS.21 &  4.2 & $ 5.9 \pm 1.4$ & $ 4.0 \pm ^{+ 1.5}_{- 1.6}$ &  0.01 & 19.3\% & 18.9\% &  1.1\arcsec \\
AzTEC\_J095950.78+022828.3 & AzTEC/COSMOS.22 &  4.3 & $ 6.6 \pm 1.5$ & $ 4.3 \pm ^{+ 1.7}_{- 1.6}$ &  0.01 & 22.6\% & 18.9\% &  0.7\arcsec \\
AzTEC\_J095931.58+023601.6 & AzTEC/COSMOS.23 &  3.9 & $ 5.7 \pm 1.5$ & $ 3.5 \pm ^{+ 1.6}_{- 1.7}$ &  0.02 & 11.8\% & 18.8\% &  0.9\arcsec \\
AzTEC\_J100038.83+023843.6 & AzTEC/COSMOS.24 &  3.8 & $ 5.6 \pm 1.5$ & $ 3.3 \pm ^{+ 1.7}_{- 1.6}$ &  0.02 & 11.2\% & 18.7\% &  1.1\arcsec \\
AzTEC\_J095950.39+024759.4 & AzTEC/COSMOS.25 &  4.2 & $ 6.0 \pm 1.4$ & $ 4.0 \pm ^{+ 1.5}_{- 1.6}$ &  0.01 & 21.7\% & 19.0\% &  1.6\arcsec \\
AzTEC\_J095959.58+023818.4 & AzTEC/COSMOS.26 &  4.0 & $ 5.9 \pm 1.5$ & $ 3.8 \pm ^{+ 1.5}_{- 1.7}$ &  0.01 & 18.3\% & 18.8\% &  1.7\arcsec \\
AzTEC\_J100039.11+024052.4 & AzTEC/COSMOS.27 &  3.9 & $ 5.8 \pm 1.5$ & $ 3.6 \pm ^{+ 1.6}_{- 1.7}$ &  0.02 & 15.0\% & 18.7\% &  0.2\arcsec \\
AzTEC\_J100004.54+023040.1 & AzTEC/COSMOS.28 &  3.9 & $ 6.0 \pm 1.5$ & $ 3.7 \pm ^{+ 1.7}_{- 1.8}$ &  0.02 & 17.3\% & 18.8\% &  0.4\arcsec \\
AzTEC\_J100026.69+023753.6 & AzTEC/COSMOS.29 &  3.9 & $ 5.8 \pm 1.5$ & $ 3.6 \pm ^{+ 1.6}_{- 1.7}$ &  0.02 & 17.3\% & 18.8\% &  1.0\arcsec \\
AzTEC\_J100003.87+023254.1 & AzTEC/COSMOS.30 &  4.1 & $ 6.1 \pm 1.5$ & $ 3.8 \pm ^{+ 1.7}_{- 1.6}$ &  0.01 & 22.5\% & 18.7\% &  0.4\arcsec \\
AzTEC\_J100034.60+023101.9 & AzTEC/COSMOS.31 &  3.9 & $ 5.9 \pm 1.5$ & $ 3.5 \pm ^{+ 1.7}_{- 1.7}$ &  0.02 & 17.2\% & 18.7\% &  0.6\arcsec \\
AzTEC\_J100020.66+022452.8 & AzTEC/COSMOS.32 &  3.6 & $ 6.1 \pm 1.7$ & $ 3.2 \pm ^{+ 1.7}_{- 2.2}$ &  0.05 & 13.6\% & 18.8\% &  0.9\arcsec \\
AzTEC\_J095911.70+023909.6 & AzTEC/COSMOS.33 &  3.9 & $ 6.0 \pm 1.6$ & $ 3.6 \pm ^{+ 1.7}_{- 1.8}$ &  0.03 & 21.8\% & 19.0\% &  0.3\arcsec \\
AzTEC\_J095946.66+023541.8 & AzTEC/COSMOS.34 &  3.6 & $ 5.3 \pm 1.5$ & $ 2.9 \pm ^{+ 1.6}_{- 1.8}$ &  0.04 & 13.5\% & 18.8\% &  0.8\arcsec \\
AzTEC\_J100026.69+023128.1 & AzTEC/COSMOS.35 &  3.8 & $ 5.8 \pm 1.5$ & $ 3.5 \pm ^{+ 1.6}_{- 1.8}$ &  0.03 & 21.4\% & 18.8\% &  0.4\arcsec \\
AzTEC\_J095914.01+023424.0 & AzTEC/COSMOS.36 &  3.7 & $ 5.6 \pm 1.5$ & $ 3.3 \pm ^{+ 1.6}_{- 1.8}$ &  0.03 & 19.0\% & 18.8\% &  1.1\arcsec \\
AzTEC\_J100016.31+024716.0 & AzTEC/COSMOS.37 &  3.5 & $ 5.3 \pm 1.5$ & $ 2.8 \pm ^{+ 1.7}_{- 1.7}$ &  0.05 & 14.7\% & 18.7\% &  0.6\arcsec \\
AzTEC\_J095951.72+024338.0 & AzTEC/COSMOS.38 &  3.6 & $ 5.1 \pm 1.4$ & $ 2.8 \pm ^{+ 1.6}_{- 1.6}$ &  0.04 & 15.2\% & 18.7\% &  0.8\arcsec \\
AzTEC\_J095958.28+023608.2 & AzTEC/COSMOS.39 &  3.6 & $ 5.3 \pm 1.5$ & $ 2.8 \pm ^{+ 1.7}_{- 1.7}$ &  0.04 & 16.3\% & 18.7\% &  0.4\arcsec \\
AzTEC\_J100031.09+022749.9 & AzTEC/COSMOS.40$^a$ &  3.3 & $ 5.4 \pm 1.6$ & -- & -- & 10.7\% & 18.9\% & -- \\
AzTEC\_J095957.33+024139.9 & AzTEC/COSMOS.41$^a$ &  3.4 & $ 4.9 \pm 1.4$ & -- & -- & 13.4\% & 18.8\% & -- \\
AzTEC\_J095930.37+023437.9 & AzTEC/COSMOS.42 &  3.5 & $ 5.2 \pm 1.5$ & $ 2.8 \pm ^{+ 1.6}_{- 1.9}$ &  0.05 & 15.1\% & 18.8\% &  1.1\arcsec \\
AzTEC\_J100023.90+022950.2 & AzTEC/COSMOS.43 &  3.5 & $ 5.4 \pm 1.5$ & $ 2.8 \pm ^{+ 1.7}_{- 1.8}$ &  0.05 & 16.3\% & 18.7\% &  0.8\arcsec \\
AzTEC\_J095920.62+023417.9 & AzTEC/COSMOS.44$^a$ &  3.3 & $ 5.0 \pm 1.5$ & -- & -- & 10.6\% & 18.8\% & -- \\
AzTEC\_J095932.26+023648.3 & AzTEC/COSMOS.45 &  3.8 & $ 5.5 \pm 1.5$ & $ 3.3 \pm ^{+ 1.6}_{- 1.7}$ &  0.03 & 24.8\% & 18.7\% &  0.7\arcsec \\
AzTEC\_J100000.79+022635.9 & AzTEC/COSMOS.46 &  3.6 & $ 5.6 \pm 1.5$ & $ 3.0 \pm ^{+ 1.7}_{- 1.9}$ &  0.04 & 19.9\% & 18.8\% &  0.3\arcsec \\
AzTEC\_J095938.54+023146.4 & AzTEC/COSMOS.47 &  3.5 & $ 5.3 \pm 1.5$ & $ 2.8 \pm ^{+ 1.7}_{- 1.8}$ &  0.05 & 17.6\% & 18.8\% &  0.7\arcsec \\
AzTEC\_J095943.85+023329.9 & AzTEC/COSMOS.48$^a$ &  3.3 & $ 5.0 \pm 1.5$ & -- & -- & 12.4\% & 18.7\% & -- \\
AzTEC\_J100039.05+024129.8 & AzTEC/COSMOS.49 &  3.7 & $ 5.5 \pm 1.5$ & $ 3.2 \pm ^{+ 1.6}_{- 1.9}$ &  0.04 & 23.7\% & 18.8\% &  0.7\arcsec \\
AzTEC\_J100012.41+022657.6 & AzTEC/COSMOS.50 &  3.6 & $ 5.6 \pm 1.5$ & $ 3.2 \pm ^{+ 1.6}_{- 2.0}$ &  0.04 & 23.4\% & 18.8\% &  0.7\arcsec \\

\hline
\multicolumn{9}{l}{The columns are as follows: (1) AzTEC source name, including RA and declination based on centroid position; (2) nickname;}\\
\multicolumn{9}{l}{(3) signal-to-noise of the detection; (4) measured $1100 \umu$m flux density and error; (5) flux density and 68 per cent confidence}\\
\multicolumn{9}{l}{interval of the deboosted flux density, including corrections for the bias to peak locations in the map; (6) probability that the}\\
\multicolumn{9}{l}{source will deboost to $S<0$ assuming the number counts prior based on all AzTEC measurements; (7,8) the relative increase}\\
\multicolumn{9}{l}{in flux and noise estimate for each source if it was detected in the previously release catalogue; (9) change in location of the}\\
\multicolumn{9}{l}{centroided source position if it was detected in both catalogs. ($^a$) indicates a source passed a significance test in the original}\\
\multicolumn{9}{l}{catalog, but not the same test in the new catalog. ($^b$) indicates a source passed a significance test in the new catalog, but not}\\
\multicolumn{9}{l}{the same test in the original catalog. In each case, an estimate for the missing quantity is made from the nearest pixel in the}\\
\multicolumn{9}{l}{map in which the test did not succeed.}
\end{tabular}
\label{tbl:cosmos}
\end{table*}

\subsection{GOODS North / JCMT}
The GOODS North field is commonly observed at many wavelengths. A revision of the catalogue presented in
\cite{PereraGOODSN2008} is shown in Table \ref{tbl:goodsn}. Similar cuts are taken at the 70 per cent coverage region (0.07 deg$^2$)
and detection significances above 3.5. The false detection rate was estimated using pure noise maps to be 13 per cent ($\sim$4-5 sources). 

\begin{table*}
\caption{The AzTEC point source catalogue for the GOODS North field.}
\begin{tabular}{|l|l|c|c|c|c|c|c|c|}
\hline \hline
Source ID & Nickname & S/N & $S_{1.1\text{mm}}$ & $S^\text{corrected}_{1.1\text{mm}}$ & $P(<0)$ & Flux & Noise & $\theta$\\
& & & [mJy] & [mJy] & & increase & increase & \\
\hline
AzTEC\_J123712.00+622210.3 & AzTEC/GN1 & 11.7 & $14.4 \pm 1.2$ & $13.5 \pm ^{+ 1.0}_{- 1.4}$ &  0.00 & 26.1\% & 25.2\% &  0.8\arcsec \\
AzTEC\_J123631.88+621709.9 & AzTEC/GN2 &  7.1 & $ 8.6 \pm 1.2$ & $ 7.4 \pm ^{+ 1.3}_{- 1.2}$ &  0.00 & 25.8\% & 25.4\% &  0.2\arcsec \\
AzTEC\_J123633.34+621408.0 & AzTEC/GN3 &  6.4 & $ 7.7 \pm 1.2$ & $ 6.5 \pm ^{+ 1.3}_{- 1.2}$ &  0.00 & 24.2\% & 25.3\% &  0.8\arcsec \\
AzTEC\_J123550.30+621044.3 & AzTEC/GN4 &  5.8 & $ 7.4 \pm 1.3$ & $ 6.1 \pm ^{+ 1.3}_{- 1.4}$ &  0.00 & 28.3\% & 25.4\% &  0.8\arcsec \\
AzTEC\_J123730.61+621256.0 & AzTEC/GN5 &  5.5 & $ 6.7 \pm 1.2$ & $ 6.5 \pm ^{+ 1.7}_{- 1.6}$ &  0.00 & 28.6\% & 25.2\% &  0.7\arcsec \\
AzTEC\_J123626.97+620605.7 & AzTEC/GN6 &  5.4 & $ 6.8 \pm 1.3$ & $ 5.4 \pm ^{+ 1.2}_{- 1.3}$ &  0.00 & 29.0\% & 25.4\% &  0.8\arcsec \\
AzTEC\_J123711.77+621330.0 & AzTEC/GN7 &  5.4 & $ 6.5 \pm 1.2$ & $ 5.4 \pm ^{+ 1.3}_{- 1.3}$ &  0.00 & 29.8\% & 25.4\% &  0.4\arcsec \\
AzTEC\_J123645.74+621442.0 & AzTEC/GN8 &  5.0 & $ 6.1 \pm 1.2$ & $ 5.2 \pm ^{+ 1.2}_{- 1.3}$ &  0.00 & 22.7\% & 25.5\% &  0.5\arcsec \\
AzTEC\_J123738.12+621735.5 & AzTEC/GN9 &  4.4 & $ 5.4 \pm 1.2$ & $ 3.8 \pm ^{+ 1.5}_{- 1.3}$ &  0.01 & 19.7\% & 25.4\% &  0.9\arcsec \\
AzTEC\_J123627.14+621217.9 & AzTEC/GN10 &  4.7 & $ 5.7 \pm 1.2$ & $ 4.6 \pm ^{+ 1.3}_{- 1.3}$ &  0.00 & 28.4\% & 25.4\% &  0.6\arcsec \\
AzTEC\_J123635.68+620706.4 & AzTEC/GN11 &  4.4 & $ 5.4 \pm 1.2$ & $ 4.2 \pm ^{+ 1.3}_{- 1.3}$ &  0.00 & 22.2\% & 25.2\% &  1.1\arcsec \\
AzTEC\_J123633.19+620617.8 & AzTEC/GN12 &  4.2 & $ 5.2 \pm 1.2$ & $ 3.8 \pm ^{+ 1.8}_{- 1.7}$ &  0.04 & 19.3\% & 25.3\% &  1.0\arcsec \\
AzTEC\_J123553.82+621344.9 & AzTEC/GN13 &  4.3 & $ 5.3 \pm 1.2$ & $ 3.8 \pm ^{+ 1.4}_{- 1.5}$ &  0.02 & 23.9\% & 25.3\% &  0.9\arcsec \\
AzTEC\_J123652.22+621224.3 & AzTEC/GN14 &  4.4 & $ 5.3 \pm 1.2$ & $ 3.8 \pm ^{+ 1.4}_{- 1.3}$ &  0.01 & 26.9\% & 25.3\% &  1.4\arcsec \\
AzTEC\_J123548.42+621528.8 & AzTEC/GN15 &  4.2 & $ 5.9 \pm 1.4$ & $ 3.7 \pm ^{+ 1.3}_{- 1.5}$ &  0.02 & 24.8\% & 25.6\% &  1.5\arcsec \\
AzTEC\_J123616.18+621517.7 & AzTEC/GN16 &  4.3 & $ 5.2 \pm 1.2$ & $ 3.8 \pm ^{+ 1.4}_{- 1.5}$ &  0.02 & 26.6\% & 25.4\% &  1.2\arcsec \\
AzTEC\_J123540.90+621436.6 & AzTEC/GN17 &  4.2 & $ 5.9 \pm 1.4$ & $ 3.8 \pm ^{+ 1.7}_{- 1.7}$ &  0.03 & 24.3\% & 24.8\% &  2.2\arcsec \\
AzTEC\_J123740.91+621221.9 & AzTEC/GN18 &  4.2 & $ 5.2 \pm 1.2$ & $ 3.9 \pm ^{+ 1.6}_{- 1.8}$ &  0.03 & 26.4\% & 25.4\% &  0.5\arcsec \\
AzTEC\_J123604.17+620701.0 & AzTEC/GN19 &  4.2 & $ 5.8 \pm 1.4$ & $ 3.6 \pm ^{+ 1.4}_{- 1.5}$ &  0.02 & 26.7\% & 25.3\% &  1.2\arcsec \\
AzTEC\_J123712.29+621037.4 & AzTEC/GN20 &  4.2 & $ 5.1 \pm 1.2$ & $ 3.5 \pm ^{+ 1.5}_{- 1.5}$ &  0.03 & 26.1\% & 25.5\% &  0.2\arcsec \\
AzTEC\_J123800.96+621613.4 & AzTEC/GN21 &  4.3 & $ 5.3 \pm 1.2$ & $ 3.8 \pm ^{+ 1.3}_{- 1.4}$ &  0.01 & 34.1\% & 25.5\% &  1.4\arcsec \\
AzTEC\_J123649.46+621210.6 & AzTEC/GN22 &  3.6 & $ 4.4 \pm 1.2$ & $ 2.8 \pm ^{+ 1.2}_{- 2.9}$ &  0.15 & 14.9\% & 25.5\% &  3.5\arcsec \\
AzTEC\_J123716.88+621731.8 & AzTEC/GN23 &  3.9 & $ 4.8 \pm 1.2$ & $ 3.1 \pm ^{+ 1.5}_{- 1.6}$ &  0.04 & 27.1\% & 25.3\% &  1.2\arcsec \\
AzTEC\_J123608.47+621441.2 & AzTEC/GN24 &  4.0 & $ 4.9 \pm 1.2$ & $ 3.1 \pm ^{+ 1.5}_{- 1.6}$ &  0.04 & 29.0\% & 25.3\% &  0.7\arcsec \\
AzTEC\_J123652.29+620503.8 & AzTEC/GN25 &  3.5 & $ 4.8 \pm 1.4$ & $ 2.4 \pm ^{+ 1.3}_{- 2.3}$ &  0.11 & 14.8\% & 25.1\% &  0.3\arcsec \\
AzTEC\_J123713.86+621825.8 & AzTEC/GN26 &  3.7 & $ 4.5 \pm 1.2$ & $ 2.8 \pm ^{+ 1.6}_{- 1.9}$ &  0.08 & 22.5\% & 25.1\% &  1.1\arcsec \\
AzTEC\_J123719.77+621221.7 & AzTEC/GN27 &  4.0 & $ 4.9 \pm 1.2$ & $ 3.4 \pm ^{+ 1.4}_{- 1.6}$ &  0.03 & 32.7\% & 25.3\% &  0.2\arcsec \\
AzTEC\_J123643.60+621935.5 & AzTEC/GN28 &  3.7 & $ 4.6 \pm 1.2$ & $ 2.8 \pm ^{+ 1.6}_{- 2.1}$ &  0.09 & 24.3\% & 25.4\% &  1.0\arcsec \\
AzTEC\_J123620.96+621912.3 & AzTEC/GN29 &  3.5 & $ 5.0 \pm 1.4$ & $ 2.4 \pm ^{+ 1.2}_{- 2.3}$ &  0.11 & 19.6\% & 25.8\% &  0.5\arcsec \\
AzTEC\_J123642.83+621718.3 & AzTEC/GN30 &  4.0 & $ 4.9 \pm 1.2$ & $ 3.2 \pm ^{+ 1.5}_{- 1.6}$ &  0.04 & 35.7\% & 25.3\% &  0.3\arcsec \\
AzTEC\_J123622.16+621611.9 & AzTEC/GN31 &  3.5 & $ 4.3 \pm 1.2$ & $ 2.5 \pm ^{+ 1.4}_{- 2.1}$ &  0.10 & 20.5\% & 25.4\% &  1.1\arcsec \\
AzTEC\_J123717.11+621357.4 & AzTEC/GN32 &  3.7 & $ 4.5 \pm 1.2$ & $ 2.8 \pm ^{+ 1.6}_{- 2.0}$ &  0.08 & 26.1\% & 25.4\% &  1.2\arcsec \\
AzTEC\_J123651.40+622023.5 & AzTEC/GN33 &  4.1 & $ 5.0 \pm 1.2$ & $ 3.5 \pm ^{+ 1.4}_{- 1.5}$ &  0.03 & 42.2\% & 25.4\% &  1.0\arcsec \\
AzTEC\_J123648.22+622105.2 & AzTEC/GN34 &  3.7 & $ 4.8 \pm 1.3$ & $ 2.7 \pm ^{+ 1.5}_{- 1.8}$ &  0.07 & 30.3\% & 25.6\% &  0.9\arcsec \\
AzTEC\_J123818.20+621430.1 & AzTEC/GN35 &  3.6 & $ 5.1 \pm 1.4$ & $ 2.7 \pm ^{+ 1.5}_{- 1.9}$ &  0.08 & 27.4\% & 25.5\% &  1.2\arcsec \\
AzTEC\_J123617.35+621547.2 & AzTEC/GN36 &  3.5 & $ 4.3 \pm 1.2$ & $ 2.7 \pm ^{+ 1.4}_{- 2.7}$ &  0.12 & 26.8\% & 25.4\% &  1.8\arcsec \\
AzTEC\_J123623.24+620331.6 & AzTEC/GN37$^b$ &  5.5 & $ 8.5 \pm 1.5$ & $ 6.5 \pm ^{+ 1.7}_{- 1.6}$ &  0.00 & 30.5\% & 25.8\% & -- \\
AzTEC\_J123645.03+622018.1 & AzTEC/GN38$^b$ &  3.8 & $ 4.7 \pm 1.2$ & $ 2.8 \pm ^{+ 1.6}_{- 1.9}$ &  0.08 & 35.3\% & 25.4\% & -- \\
AzTEC\_J123546.21+621152.2 & AzTEC/GN39$^b$ &  3.7 & $ 4.7 \pm 1.3$ & $ 2.8 \pm ^{+ 1.6}_{- 2.0}$ &  0.08 & 42.5\% & 25.5\% & -- \\
AzTEC\_J123629.32+620257.8 & AzTEC/GN40$^b$ &  3.6 & $ 5.8 \pm 1.6$ & $ 2.8 \pm ^{+ 1.2}_{- 2.9}$ &  0.15 & 20.4\% & 25.2\% & -- \\

\hline
\multicolumn{9}{l}{Columns are as described in Table \ref{tbl:cosmos}.}
\end{tabular}
\label{tbl:goodsn}
\end{table*}

\subsection{Lockman Hole / JCMT}
The Lockman Hole survey \citep{AustermannSHADES2010} was undertaken by the AzTEC instrument in 2005 while it was 
installed at the JCMT and formed part of the 1.1mm follow-up to the SCUBA/SHADES project. A revised catalogue is shown in 
Table \ref{tbl:lh}. The 50 per cent coverage region (0.31 deg$^2$) was selected as a spatial cut, but a different significance cut was used. 
Deboosting can also be used as a proxy for whether a source is likely to be real. If the likelihood of deboosting to 0 flux is significant,
then that source can be excluded from the catalogue. Only sources with less than 10 per cent likelihood of deboosting to 0 flux are taken.
The false detection rate using these cuts was estimated by the technique described in \cite{PereraGOODSN2008,ScottGOODSS2010}.
In this technique, false detection rates are estimated by fully simulating maps using noise estimates and a signal estimate using the
number counts used in deboosting measured fluxes. Using this technique, the false detection rate was estimated to be 20 per cent
($\sim$20 sources).

\begin{table*}
\caption{The AzTEC point source catalogue for the Lockman Hole field.}
\begin{tabular}{|l|l|c|c|c|c|c|c|c|}
\hline \hline
Source ID & Nickname & S/N & $S_{1.1\text{mm}}$ & $S^\text{corrected}_{1.1\text{mm}}$ & $P(<0)$ & Flux & Noise & $\theta$\\
& & & [mJy] & [mJy] & & increase & increase & \\
\hline
AzTEC\_J105201.98+574049.2 & AzTEC/LH1 &  8.0 & $ 8.8 \pm 1.1$ & $ 7.8 \pm ^{+ 1.2}_{- 1.1}$ &  0.00 & 19.6\% & 22.4\% &  0.4\arcsec \\
AzTEC\_J105206.17+573623.1 & AzTEC/LH2 &  8.0 & $ 8.6 \pm 1.1$ & $ 7.6 \pm ^{+ 1.2}_{- 1.0}$ &  0.00 & 19.0\% & 22.4\% &  1.2\arcsec \\
AzTEC\_J105257.12+572104.5 & AzTEC/LH3 &  7.4 & $ 8.8 \pm 1.2$ & $ 7.6 \pm ^{+ 1.3}_{- 1.1}$ &  0.00 & 21.0\% & 22.4\% &  1.0\arcsec \\
AzTEC\_J105044.49+573319.3 & AzTEC/LH4 &  6.7 & $ 7.5 \pm 1.1$ & $ 6.5 \pm ^{+ 1.1}_{- 1.2}$ &  0.00 & 22.5\% & 22.5\% &  1.0\arcsec \\
AzTEC\_J105403.64+572552.9 & AzTEC/LH5 &  6.6 & $ 7.5 \pm 1.1$ & $ 6.5 \pm ^{+ 1.0}_{- 1.3}$ &  0.00 & 26.6\% & 22.5\% &  0.5\arcsec \\
AzTEC\_J105241.84+573551.1 & AzTEC/LH6 &  6.1 & $ 6.7 \pm 1.1$ & $ 5.7 \pm ^{+ 1.1}_{- 1.1}$ &  0.00 & 20.6\% & 22.5\% &  1.3\arcsec \\
AzTEC\_J105203.95+572659.4 & AzTEC/LH7 &  6.1 & $ 7.1 \pm 1.2$ & $ 6.0 \pm ^{+ 1.1}_{- 1.3}$ &  0.00 & 24.6\% & 22.5\% &  0.4\arcsec \\
AzTEC\_J105201.02+572443.2 & AzTEC/LH8 &  5.9 & $ 6.8 \pm 1.2$ & $ 5.5 \pm ^{+ 1.3}_{- 1.1}$ &  0.00 & 20.1\% & 22.3\% &  0.5\arcsec \\
AzTEC\_J105214.15+573326.6 & AzTEC/LH9 &  5.7 & $ 6.1 \pm 1.1$ & $ 5.1 \pm ^{+ 1.1}_{- 1.1}$ &  0.00 & 23.3\% & 22.4\% &  0.2\arcsec \\
AzTEC\_J105406.43+573310.6 & AzTEC/LH10 &  5.3 & $ 6.1 \pm 1.1$ & $ 4.8 \pm ^{+ 1.3}_{- 1.1}$ &  0.00 & 18.9\% & 22.4\% &  1.0\arcsec \\
AzTEC\_J105130.39+573807.1 & AzTEC/LH11 &  5.6 & $ 6.1 \pm 1.1$ & $ 5.0 \pm ^{+ 1.2}_{- 1.1}$ &  0.00 & 27.4\% & 22.3\% &  1.0\arcsec \\
AzTEC\_J105217.18+573502.8 & AzTEC/LH12 &  5.2 & $ 5.6 \pm 1.1$ & $ 4.4 \pm ^{+ 1.2}_{- 1.1}$ &  0.00 & 18.6\% & 22.4\% &  0.6\arcsec \\
AzTEC\_J105140.73+574323.2 & AzTEC/LH13 &  5.1 & $ 6.3 \pm 1.2$ & $ 4.8 \pm ^{+ 1.4}_{- 1.3}$ &  0.00 & 18.1\% & 22.0\% &  2.0\arcsec \\
AzTEC\_J105220.16+573956.6 & AzTEC/LH14 &  5.2 & $ 5.7 \pm 1.1$ & $ 4.6 \pm ^{+ 1.1}_{- 1.2}$ &  0.00 & 23.9\% & 22.5\% &  1.6\arcsec \\
AzTEC\_J105256.38+574227.8 & AzTEC/LH15 &  5.0 & $ 5.7 \pm 1.1$ & $ 4.4 \pm ^{+ 1.2}_{- 1.2}$ &  0.00 & 19.4\% & 22.4\% &  0.6\arcsec \\
AzTEC\_J105341.56+573215.8 & AzTEC/LH16 &  4.9 & $ 5.5 \pm 1.1$ & $ 4.2 \pm ^{+ 1.2}_{- 1.2}$ &  0.00 & 16.9\% & 22.4\% &  1.2\arcsec \\
AzTEC\_J105319.57+572105.0 & AzTEC/LH17 &  5.0 & $ 5.7 \pm 1.2$ & $ 4.4 \pm ^{+ 1.2}_{- 1.2}$ &  0.00 & 20.4\% & 22.4\% &  0.1\arcsec \\
AzTEC\_J105225.16+573836.5 & AzTEC/LH18 &  4.8 & $ 5.1 \pm 1.1$ & $ 3.9 \pm ^{+ 1.2}_{- 1.1}$ &  0.00 & 21.0\% & 22.4\% &  1.0\arcsec \\
AzTEC\_J105129.62+573650.6 & AzTEC/LH19 &  4.7 & $ 5.1 \pm 1.1$ & $ 3.9 \pm ^{+ 1.2}_{- 1.2}$ &  0.00 & 20.7\% & 22.4\% &  1.3\arcsec \\
AzTEC\_J105345.54+571647.8 & AzTEC/LH20 &  5.1 & $ 6.3 \pm 1.2$ & $ 5.0 \pm ^{+ 1.2}_{- 1.4}$ &  0.00 & 34.2\% & 22.4\% &  0.3\arcsec \\
AzTEC\_J105131.43+573133.4 & AzTEC/LH21 &  4.4 & $ 4.8 \pm 1.1$ & $ 3.5 \pm ^{+ 1.2}_{- 1.2}$ &  0.01 & 16.2\% & 22.4\% &  1.7\arcsec \\
AzTEC\_J105256.43+572356.2 & AzTEC/LH22 &  4.9 & $ 5.8 \pm 1.2$ & $ 4.4 \pm ^{+ 1.3}_{- 1.2}$ &  0.00 & 29.5\% & 22.6\% &  0.3\arcsec \\
AzTEC\_J105321.99+571718.0 & AzTEC/LH23 &  4.8 & $ 5.7 \pm 1.2$ & $ 4.4 \pm ^{+ 1.2}_{- 1.4}$ &  0.00 & 29.6\% & 22.4\% &  2.1\arcsec \\
AzTEC\_J105238.54+572436.8 & AzTEC/LH24 &  4.7 & $ 5.4 \pm 1.2$ & $ 4.1 \pm ^{+ 1.3}_{- 1.3}$ &  0.00 & 27.9\% & 22.4\% &  1.1\arcsec \\
AzTEC\_J105107.01+573442.1 & AzTEC/LH25 &  4.7 & $ 5.2 \pm 1.1$ & $ 3.9 \pm ^{+ 1.2}_{- 1.2}$ &  0.00 & 31.7\% & 22.5\% &  0.3\arcsec \\
AzTEC\_J105059.75+571637.6 & AzTEC/LH26 &  4.4 & $ 5.6 \pm 1.3$ & $ 3.9 \pm ^{+ 1.4}_{- 1.5}$ &  0.01 & 23.2\% & 22.4\% &  1.0\arcsec \\
AzTEC\_J105218.61+571853.5 & AzTEC/LH27 &  4.4 & $ 5.2 \pm 1.2$ & $ 3.7 \pm ^{+ 1.4}_{- 1.4}$ &  0.01 & 23.1\% & 22.5\% &  1.2\arcsec \\
AzTEC\_J105045.28+573649.2 & AzTEC/LH28 &  4.4 & $ 5.1 \pm 1.2$ & $ 3.7 \pm ^{+ 1.3}_{- 1.4}$ &  0.01 & 25.4\% & 22.0\% &  0.4\arcsec \\
AzTEC\_J105123.33+572200.6 & AzTEC/LH29 &  4.3 & $ 5.0 \pm 1.2$ & $ 3.5 \pm ^{+ 1.3}_{- 1.3}$ &  0.01 & 23.8\% & 22.5\% &  0.6\arcsec \\
AzTEC\_J105238.21+573002.6 & AzTEC/LH30 &  4.2 & $ 4.6 \pm 1.1$ & $ 3.3 \pm ^{+ 1.2}_{- 1.3}$ &  0.02 & 20.8\% & 22.5\% &  0.3\arcsec \\
AzTEC\_J105425.19+573707.7 & AzTEC/LH31 &  4.1 & $ 6.2 \pm 1.5$ & $ 3.9 \pm ^{+ 1.8}_{- 2.1}$ &  0.05 & 19.7\% & 22.0\% &  0.3\arcsec \\
AzTEC\_J105041.20+572129.6 & AzTEC/LH32 &  4.1 & $ 5.0 \pm 1.2$ & $ 3.3 \pm ^{+ 1.5}_{- 1.5}$ &  0.03 & 19.6\% & 22.5\% &  1.0\arcsec \\
AzTEC\_J105246.40+573120.8 & AzTEC/LH33 &  4.2 & $ 4.6 \pm 1.1$ & $ 3.3 \pm ^{+ 1.2}_{- 1.3}$ &  0.02 & 22.2\% & 22.7\% &  3.3\arcsec \\
AzTEC\_J105238.37+572324.4 & AzTEC/LH34 &  4.1 & $ 4.8 \pm 1.2$ & $ 3.3 \pm ^{+ 1.4}_{- 1.4}$ &  0.03 & 21.9\% & 22.4\% &  0.7\arcsec \\
AzTEC\_J105355.86+572953.9 & AzTEC/LH35 &  4.0 & $ 4.5 \pm 1.1$ & $ 3.1 \pm ^{+ 1.3}_{- 1.4}$ &  0.03 & 20.2\% & 22.5\% &  1.4\arcsec \\
AzTEC\_J105349.54+571604.4 & AzTEC/LH36 &  4.1 & $ 5.3 \pm 1.3$ & $ 3.5 \pm ^{+ 1.5}_{- 1.7}$ &  0.04 & 21.4\% & 22.3\% &  0.8\arcsec \\
AzTEC\_J105152.68+571335.1 & AzTEC/LH37 &  4.0 & $ 5.0 \pm 1.2$ & $ 3.3 \pm ^{+ 1.5}_{- 1.7}$ &  0.04 & 20.0\% & 22.4\% &  1.6\arcsec \\
AzTEC\_J105116.29+573210.5 & AzTEC/LH38 &  4.1 & $ 4.4 \pm 1.1$ & $ 3.1 \pm ^{+ 1.3}_{- 1.2}$ &  0.02 & 25.4\% & 22.4\% &  0.6\arcsec \\
AzTEC\_J105212.28+571553.0 & AzTEC/LH39 &  4.1 & $ 5.0 \pm 1.2$ & $ 3.3 \pm ^{+ 1.5}_{- 1.5}$ &  0.03 & 24.5\% & 22.3\% &  1.0\arcsec \\
AzTEC\_J105226.59+573355.2 & AzTEC/LH40 &  3.9 & $ 4.2 \pm 1.1$ & $ 2.8 \pm ^{+ 1.3}_{- 1.4}$ &  0.04 & 18.9\% & 22.5\% &  0.4\arcsec \\
AzTEC\_J105116.32+574026.9 & AzTEC/LH41 &  4.1 & $ 4.7 \pm 1.1$ & $ 3.3 \pm ^{+ 1.3}_{- 1.5}$ &  0.03 & 26.2\% & 22.3\% &  0.3\arcsec \\
AzTEC\_J105058.32+571843.8 & AzTEC/LH42 &  3.9 & $ 4.7 \pm 1.2$ & $ 3.1 \pm ^{+ 1.5}_{- 1.6}$ &  0.05 & 21.6\% & 22.4\% &  0.6\arcsec \\
AzTEC\_J105153.10+572123.2 & AzTEC/LH43 &  3.9 & $ 4.6 \pm 1.2$ & $ 3.1 \pm ^{+ 1.4}_{- 1.6}$ &  0.04 & 21.8\% & 22.4\% &  0.9\arcsec \\
AzTEC\_J105241.76+573404.6 & AzTEC/LH44 &  3.7 & $ 4.0 \pm 1.1$ & $ 2.6 \pm ^{+ 1.4}_{- 1.5}$ &  0.06 & 15.7\% & 22.3\% &  0.5\arcsec \\
AzTEC\_J105154.75+573823.3 & AzTEC/LH45 &  4.0 & $ 4.4 \pm 1.1$ & $ 3.1 \pm ^{+ 1.2}_{- 1.4}$ &  0.03 & 25.4\% & 22.4\% &  1.5\arcsec \\
AzTEC\_J105210.62+571432.8 & AzTEC/LH46 &  3.9 & $ 4.8 \pm 1.2$ & $ 3.1 \pm ^{+ 1.5}_{- 1.8}$ &  0.05 & 20.8\% & 22.4\% &  0.2\arcsec \\
AzTEC\_J105307.00+573031.9 & AzTEC/LH47 &  3.9 & $ 4.4 \pm 1.1$ & $ 2.9 \pm ^{+ 1.3}_{- 1.5}$ &  0.04 & 22.9\% & 22.4\% &  0.7\arcsec \\
AzTEC\_J105431.35+572543.2 & AzTEC/LH48 &  3.9 & $ 4.9 \pm 1.3$ & $ 3.1 \pm ^{+ 1.5}_{- 1.8}$ &  0.06 & 20.7\% & 22.5\% &  0.9\arcsec \\
AzTEC\_J105340.43+572754.0 & AzTEC/LH49 &  3.7 & $ 4.2 \pm 1.1$ & $ 2.6 \pm ^{+ 1.4}_{- 1.7}$ &  0.06 & 16.5\% & 22.5\% &  0.4\arcsec \\
AzTEC\_J105205.46+572916.6 & AzTEC/LH50 &  3.8 & $ 4.2 \pm 1.1$ & $ 2.6 \pm ^{+ 1.4}_{- 1.5}$ &  0.05 & 19.5\% & 22.3\% &  0.8\arcsec \\
AzTEC\_J105035.96+573332.4 & AzTEC/LH51 &  3.9 & $ 4.6 \pm 1.2$ & $ 3.1 \pm ^{+ 1.4}_{- 1.6}$ &  0.04 & 24.1\% & 22.3\% &  0.5\arcsec \\
AzTEC\_J105206.71+574538.3 & AzTEC/LH52 &  3.8 & $ 5.2 \pm 1.4$ & $ 3.1 \pm ^{+ 1.8}_{- 2.1}$ &  0.07 & 21.8\% & 22.8\% &  1.6\arcsec \\
AzTEC\_J105435.19+572715.4 & AzTEC/LH53 &  3.8 & $ 4.8 \pm 1.3$ & $ 2.9 \pm ^{+ 1.6}_{- 2.0}$ &  0.07 & 19.7\% & 22.4\% &  1.2\arcsec \\
AzTEC\_J105351.49+572649.3 & AzTEC/LH54 &  3.9 & $ 4.4 \pm 1.1$ & $ 2.9 \pm ^{+ 1.3}_{- 1.5}$ &  0.04 & 25.0\% & 22.4\% &  1.4\arcsec \\
AzTEC\_J105153.94+571034.7 & AzTEC/LH55 &  3.9 & $ 5.7 \pm 1.5$ & $ 3.3 \pm ^{+ 1.9}_{- 2.3}$ &  0.08 & 24.5\% & 22.3\% &  1.1\arcsec \\

\hline
\multicolumn{9}{l}{Columns are as described in Table \ref{tbl:cosmos}.}
\end{tabular}
\label{tbl:lh}
\end{table*}

\begin{table*}
\contcaption{}
\begin{tabular}{|l|l|c|c|c|c|c|c|c|}
\hline \hline
Source ID & Nickname & S/N & $S_{1.1\text{mm}}$ & $S^\text{corrected}_{1.1\text{mm}}$ & $P(<0)$ & Flux & Noise & $\theta$\\
& & & [mJy] & [mJy] & & increase & increase & \\
\hline
AzTEC\_J105203.76+572523.1 & AzTEC/LH56 &  3.8 & $ 4.4 \pm 1.2$ & $ 2.8 \pm ^{+ 1.5}_{- 1.6}$ &  0.05 & 23.3\% & 22.5\% &  2.2\arcsec \\
AzTEC\_J105251.44+572610.0 & AzTEC/LH57 &  3.7 & $ 4.2 \pm 1.2$ & $ 2.6 \pm ^{+ 1.4}_{- 1.8}$ &  0.08 & 17.7\% & 22.5\% &  1.4\arcsec \\
AzTEC\_J105243.69+574042.8 & AzTEC/LH58 &  3.9 & $ 4.3 \pm 1.1$ & $ 2.8 \pm ^{+ 1.4}_{- 1.4}$ &  0.04 & 26.3\% & 22.5\% &  0.5\arcsec \\
AzTEC\_J105044.99+573031.3 & AzTEC/LH59 &  3.7 & $ 4.1 \pm 1.1$ & $ 2.6 \pm ^{+ 1.4}_{- 1.6}$ &  0.06 & 19.9\% & 22.4\% &  1.2\arcsec \\
AzTEC\_J105345.57+572645.7 & AzTEC/LH60 &  3.9 & $ 4.4 \pm 1.1$ & $ 2.9 \pm ^{+ 1.3}_{- 1.5}$ &  0.04 & 26.9\% & 22.5\% &  1.0\arcsec \\
AzTEC\_J105257.10+572249.6 & AzTEC/LH61 &  3.8 & $ 4.5 \pm 1.2$ & $ 2.8 \pm ^{+ 1.6}_{- 1.6}$ &  0.05 & 24.0\% & 22.4\% &  1.2\arcsec \\
AzTEC\_J105211.46+573511.3 & AzTEC/LH62 &  3.7 & $ 4.0 \pm 1.1$ & $ 2.6 \pm ^{+ 1.3}_{- 1.6}$ &  0.06 & 21.4\% & 22.4\% &  1.0\arcsec \\
AzTEC\_J105406.19+572042.7 & AzTEC/LH63 &  3.8 & $ 4.6 \pm 1.2$ & $ 2.8 \pm ^{+ 1.6}_{- 1.8}$ &  0.07 & 23.5\% & 22.5\% &  0.9\arcsec \\
AzTEC\_J105310.85+573436.0 & AzTEC/LH64 &  3.8 & $ 4.2 \pm 1.1$ & $ 2.6 \pm ^{+ 1.4}_{- 1.4}$ &  0.05 & 24.9\% & 22.4\% &  0.5\arcsec \\
AzTEC\_J105258.33+573935.3 & AzTEC/LH65 &  3.7 & $ 4.1 \pm 1.1$ & $ 2.6 \pm ^{+ 1.4}_{- 1.7}$ &  0.07 & 20.7\% & 22.3\% &  1.1\arcsec \\
AzTEC\_J105351.71+573052.2 & AzTEC/LH66 &  3.6 & $ 4.0 \pm 1.1$ & $ 2.3 \pm ^{+ 1.4}_{- 1.8}$ &  0.09 & 17.6\% & 22.3\% &  6.4\arcsec \\
AzTEC\_J105045.36+572925.1 & AzTEC/LH67 &  3.6 & $ 4.0 \pm 1.1$ & $ 2.5 \pm ^{+ 1.4}_{- 1.7}$ &  0.08 & 20.0\% & 22.4\% &  0.4\arcsec \\
AzTEC\_J105326.00+572247.2 & AzTEC/LH68 &  3.8 & $ 4.4 \pm 1.1$ & $ 2.8 \pm ^{+ 1.4}_{- 1.6}$ &  0.05 & 26.2\% & 22.6\% &  0.6\arcsec \\
AzTEC\_J105059.84+573246.0 & AzTEC/LH69 &  3.9 & $ 4.2 \pm 1.1$ & $ 2.8 \pm ^{+ 1.3}_{- 1.5}$ &  0.04 & 27.8\% & 22.4\% &  0.8\arcsec \\
AzTEC\_J105121.56+573332.9 & AzTEC/LH70 &  3.7 & $ 4.0 \pm 1.1$ & $ 2.6 \pm ^{+ 1.3}_{- 1.5}$ &  0.06 & 23.7\% & 22.4\% &  1.0\arcsec \\
AzTEC\_J105406.84+572959.2 & AzTEC/LH71 &  3.6 & $ 4.1 \pm 1.1$ & $ 2.5 \pm ^{+ 1.4}_{- 1.8}$ &  0.08 & 20.4\% & 22.3\% &  1.9\arcsec \\
AzTEC\_J105132.65+574022.7 & AzTEC/LH72 &  3.7 & $ 4.1 \pm 1.1$ & $ 2.6 \pm ^{+ 1.4}_{- 1.7}$ &  0.07 & 22.0\% & 22.5\% &  0.6\arcsec \\
AzTEC\_J105157.02+574057.0 & AzTEC/LH73 &  3.6 & $ 3.9 \pm 1.1$ & $ 2.3 \pm ^{+ 1.4}_{- 1.7}$ &  0.08 & 18.7\% & 22.3\% &  0.4\arcsec \\
AzTEC\_J105246.39+571742.4 & AzTEC/LH74$^a$ &  3.2 & $ 4.0 \pm 1.2$ & -- & -- &  9.5\% & 22.5\% & -- \\
AzTEC\_J105309.81+571659.7 & AzTEC/LH75$^a$ &  3.4 & $ 4.2 \pm 1.2$ & -- & -- & 16.5\% & 22.5\% & -- \\
AzTEC\_J105228.38+573258.4 & AzTEC/LH76 &  3.8 & $ 4.0 \pm 1.1$ & $ 2.6 \pm ^{+ 1.3}_{- 1.5}$ &  0.05 & 26.4\% & 22.4\% &  0.5\arcsec \\
AzTEC\_J105148.09+574123.0 & AzTEC/LH77 &  3.9 & $ 4.3 \pm 1.1$ & $ 2.8 \pm ^{+ 1.3}_{- 1.5}$ &  0.04 & 29.1\% & 22.4\% &  1.3\arcsec \\
AzTEC\_J105349.87+573352.0 & AzTEC/LH78$^a$ &  3.2 & $ 3.6 \pm 1.1$ & -- & -- &  7.6\% & 22.3\% & -- \\
AzTEC\_J105232.55+571540.8 & AzTEC/LH79 &  3.8 & $ 4.5 \pm 1.2$ & $ 2.8 \pm ^{+ 1.5}_{- 1.8}$ &  0.07 & 25.6\% & 22.5\% &  0.4\arcsec \\
AzTEC\_J105418.69+573448.0 & AzTEC/LH80 &  3.8 & $ 4.7 \pm 1.2$ & $ 2.9 \pm ^{+ 1.5}_{- 1.8}$ &  0.06 & 28.0\% & 22.8\% &  1.5\arcsec \\
AzTEC\_J105321.62+572307.4 & AzTEC/LH81 &  3.7 & $ 4.2 \pm 1.1$ & $ 2.6 \pm ^{+ 1.4}_{- 1.8}$ &  0.07 & 24.3\% & 22.6\% &  0.4\arcsec \\
AzTEC\_J105136.87+573758.7 & AzTEC/LH82 &  3.7 & $ 4.0 \pm 1.1$ & $ 2.5 \pm ^{+ 1.3}_{- 1.7}$ &  0.07 & 23.7\% & 22.4\% &  0.3\arcsec \\
AzTEC\_J105343.83+572544.3 & AzTEC/LH83 &  3.6 & $ 4.1 \pm 1.1$ & $ 2.5 \pm ^{+ 1.4}_{- 1.8}$ &  0.08 & 24.1\% & 22.3\% &  1.2\arcsec \\
AzTEC\_J105230.64+572208.9 & AzTEC/LH84 &  3.7 & $ 4.4 \pm 1.2$ & $ 2.6 \pm ^{+ 1.5}_{- 1.7}$ &  0.07 & 27.4\% & 22.6\% &  2.3\arcsec \\
AzTEC\_J105036.80+573228.7 & AzTEC/LH85 &  3.8 & $ 4.4 \pm 1.1$ & $ 2.8 \pm ^{+ 1.4}_{- 1.6}$ &  0.06 & 30.4\% & 22.6\% &  0.9\arcsec \\
AzTEC\_J105037.16+572845.3 & AzTEC/LH86$^a$ &  3.4 & $ 3.9 \pm 1.1$ & -- & -- & 20.1\% & 22.4\% & -- \\
AzTEC\_J105044.88+573421.4 & AzTEC/LH87$^b$ &  3.8 & $ 4.2 \pm 1.1$ & $ 2.6 \pm ^{+ 1.4}_{- 1.6}$ &  0.06 & 31.9\% & 22.5\% & -- \\
AzTEC\_J105202.03+571445.1 & AzTEC/LH88$^b$ &  3.7 & $ 4.5 \pm 1.2$ & $ 2.6 \pm ^{+ 1.5}_{- 2.0}$ &  0.09 & 36.8\% & 22.5\% & -- \\
AzTEC\_J105205.94+574203.3 & AzTEC/LH89$^b$ &  3.7 & $ 4.1 \pm 1.1$ & $ 2.5 \pm ^{+ 1.4}_{- 1.7}$ &  0.07 & 31.9\% & 22.4\% & -- \\
AzTEC\_J105158.34+574336.5 & AzTEC/LH90$^b$ &  3.6 & $ 4.3 \pm 1.2$ & $ 2.6 \pm ^{+ 1.5}_{- 1.9}$ &  0.08 & 39.9\% & 22.5\% & -- \\
AzTEC\_J105313.22+572127.9 & AzTEC/LH91$^b$ &  3.6 & $ 4.2 \pm 1.2$ & $ 2.5 \pm ^{+ 1.4}_{- 1.9}$ &  0.09 & 34.6\% & 22.5\% & -- \\
AzTEC\_J105107.70+572614.5 & AzTEC/LH92$^b$ &  3.6 & $ 4.0 \pm 1.1$ & $ 2.4 \pm ^{+ 1.4}_{- 1.8}$ &  0.08 & 29.4\% & 22.6\% & -- \\
AzTEC\_J105129.86+572502.3 & AzTEC/LH93$^b$ &  3.5 & $ 4.0 \pm 1.1$ & $ 2.3 \pm ^{+ 1.3}_{- 2.0}$ &  0.10 & 26.8\% & 22.5\% & -- \\
AzTEC\_J105147.00+573732.9 & AzTEC/LH94$^b$ &  3.5 & $ 3.8 \pm 1.1$ & $ 2.2 \pm ^{+ 1.3}_{- 1.8}$ &  0.09 & 24.5\% & 22.4\% & -- \\
AzTEC\_J105336.62+573222.4 & AzTEC/LH95$^b$ &  3.5 & $ 3.9 \pm 1.1$ & $ 2.2 \pm ^{+ 1.3}_{- 1.9}$ &  0.10 & 32.3\% & 22.5\% & -- \\

\hline
\multicolumn{9}{l}{Columns are as described in Table \ref{tbl:cosmos}.}
\end{tabular}
\label{tbl:lh}
\end{table*}

\subsection{Subaru XMM-Newton Deep Field / JCMT}
The Subaru XMM-Newton Deep Field (SXDF) was also surveyed as part of the SHADES followup project. A revised 
catalogue is shown in Table \ref{tbl:sxdf}. The same cuts as in the Lockman Hole field are applied with a resulting survey area
of 0.37 deg$^2$. Using the simulated map technique, the false detection rate was estimated to be 25 per cent ($\sim$16 sources).
The near doubling of the number of detected sources under the revised kernel is an artefact of using the deboosted flux values
and a different number counts model as a catalogue cut in a population of sources whose numbers grow rapidly with
declining flux.

\begin{table*}
\caption{The AzTEC point source catalogue for the Subaru XMM-Newton Deep Field.}
\begin{tabular}{|l|l|c|c|c|c|c|c|c|}
\hline \hline
Source ID & Nickname & S/N & $S_{1.1\text{mm}}$ & $S^\text{corrected}_{1.1\text{mm}}$ & $P(<0)$ & Flux & Noise & $\theta$\\
& & & [mJy] & [mJy] & & increase & increase & \\
\hline
AzTEC\_J021738.58-043331.1 & AzTEC/SXDF1 &  5.0 & $ 8.4 \pm 1.7$ & $ 6.0 \pm ^{+ 1.8}_{- 1.7}$ &  0.00 & 13.2\% & 17.1\% &  1.0\arcsec \\
AzTEC\_J021745.80-044747.3 & AzTEC/SXDF2 &  4.6 & $ 6.0 \pm 1.3$ & $ 4.4 \pm ^{+ 1.4}_{- 1.4}$ &  0.00 & 10.8\% & 17.1\% &  0.4\arcsec \\
AzTEC\_J021754.90-044724.3 & AzTEC/SXDF3 &  4.8 & $ 6.2 \pm 1.3$ & $ 4.7 \pm ^{+ 1.3}_{- 1.4}$ &  0.00 & 16.3\% & 17.2\% &  0.5\arcsec \\
AzTEC\_J021831.18-043912.7 & AzTEC/SXDF4 &  4.5 & $ 7.5 \pm 1.7$ & $ 5.0 \pm ^{+ 1.8}_{- 1.8}$ &  0.01 &  9.6\% & 17.0\% &  1.2\arcsec \\
AzTEC\_J021742.05-045626.5 & AzTEC/SXDF5 &  4.8 & $ 6.1 \pm 1.3$ & $ 4.6 \pm ^{+ 1.3}_{- 1.4}$ &  0.00 & 19.3\% & 17.0\% &  0.4\arcsec \\
AzTEC\_J021842.44-045931.3 & AzTEC/SXDF6 &  4.6 & $ 6.8 \pm 1.5$ & $ 4.8 \pm ^{+ 1.6}_{- 1.5}$ &  0.00 & 16.4\% & 17.0\% &  1.7\arcsec \\
AzTEC\_J021655.79-044532.2 & AzTEC/SXDF7 &  4.7 & $ 7.6 \pm 1.6$ & $ 5.2 \pm ^{+ 1.8}_{- 1.7}$ &  0.00 & 20.5\% & 17.1\% &  1.9\arcsec \\
AzTEC\_J021742.08-043135.0 & AzTEC/SXDF8 &  4.3 & $ 7.2 \pm 1.7$ & $ 4.7 \pm ^{+ 1.8}_{- 1.9}$ &  0.01 & 12.2\% & 17.0\% &  0.5\arcsec \\
AzTEC\_J021823.16-051137.5 & AzTEC/SXDF9 &  4.2 & $ 7.8 \pm 1.9$ & $ 4.6 \pm ^{+ 2.0}_{- 2.3}$ &  0.02 & 12.9\% & 17.3\% &  0.5\arcsec \\
AzTEC\_J021816.17-045512.9 & AzTEC/SXDF10 &  4.3 & $ 5.5 \pm 1.3$ & $ 3.9 \pm ^{+ 1.3}_{- 1.4}$ &  0.00 & 17.2\% & 17.1\% &  1.7\arcsec \\
AzTEC\_J021708.19-045617.0 & AzTEC/SXDF11 &  4.3 & $ 6.6 \pm 1.5$ & $ 4.4 \pm ^{+ 1.6}_{- 1.7}$ &  0.01 & 19.2\% & 16.8\% &  1.2\arcsec \\
AzTEC\_J021708.07-044257.0 & AzTEC/SXDF12 &  4.1 & $ 6.6 \pm 1.6$ & $ 4.0 \pm ^{+ 1.8}_{- 1.9}$ &  0.02 & 13.3\% & 17.0\% &  0.8\arcsec \\
AzTEC\_J021829.17-045448.5 & AzTEC/SXDF13 &  4.4 & $ 5.5 \pm 1.2$ & $ 4.0 \pm ^{+ 1.3}_{- 1.4}$ &  0.00 & 23.8\% & 16.9\% &  0.6\arcsec \\
AzTEC\_J021740.60-044609.2 & AzTEC/SXDF14 &  4.0 & $ 5.4 \pm 1.4$ & $ 3.6 \pm ^{+ 1.4}_{- 1.6}$ &  0.01 & 12.9\% & 17.0\% &  1.1\arcsec \\
AzTEC\_J021754.68-044417.1 & AzTEC/SXDF15 &  4.3 & $ 5.8 \pm 1.4$ & $ 4.0 \pm ^{+ 1.5}_{- 1.5}$ &  0.01 & 21.9\% & 17.1\% &  1.2\arcsec \\
AzTEC\_J021716.25-045807.3 & AzTEC/SXDF16 &  4.1 & $ 5.8 \pm 1.4$ & $ 3.8 \pm ^{+ 1.6}_{- 1.6}$ &  0.01 & 16.0\% & 17.0\% &  0.7\arcsec \\
AzTEC\_J021711.57-044315.2 & AzTEC/SXDF17 &  4.0 & $ 6.4 \pm 1.6$ & $ 4.0 \pm ^{+ 1.7}_{- 1.9}$ &  0.02 & 15.3\% & 17.0\% &  0.9\arcsec \\
AzTEC\_J021724.44-043144.5 & AzTEC/SXDF18 &  4.1 & $ 7.1 \pm 1.8$ & $ 4.2 \pm ^{+ 1.9}_{- 2.2}$ &  0.03 & 16.7\% & 17.1\% &  0.8\arcsec \\
AzTEC\_J021906.23-045334.4 & AzTEC/SXDF19 &  4.3 & $ 8.2 \pm 1.9$ & $ 5.0 \pm ^{+ 2.1}_{- 2.2}$ &  0.02 & 26.1\% & 17.3\% &  0.6\arcsec \\
AzTEC\_J021742.09-050722.8 & AzTEC/SXDF20 &  4.1 & $ 6.8 \pm 1.7$ & $ 4.1 \pm ^{+ 1.9}_{- 2.0}$ &  0.02 & 19.1\% & 16.9\% &  0.7\arcsec \\
AzTEC\_J021809.80-050444.7 & AzTEC/SXDF21 &  4.0 & $ 5.9 \pm 1.5$ & $ 3.6 \pm ^{+ 1.7}_{- 1.7}$ &  0.02 & 16.2\% & 17.2\% &  0.3\arcsec \\
AzTEC\_J021827.94-045319.0 & AzTEC/SXDF22 &  4.2 & $ 5.3 \pm 1.3$ & $ 3.7 \pm ^{+ 1.4}_{- 1.4}$ &  0.01 & 25.5\% & 17.0\% &  1.0\arcsec \\
AzTEC\_J021820.20-045738.7 & AzTEC/SXDF23 &  3.7 & $ 4.8 \pm 1.3$ & $ 3.0 \pm ^{+ 1.3}_{- 1.5}$ &  0.02 & 12.4\% & 17.2\% &  0.9\arcsec \\
AzTEC\_J021843.73-043859.3 & AzTEC/SXDF24 &  3.8 & $ 6.5 \pm 1.7$ & $ 3.6 \pm ^{+ 2.0}_{- 2.7}$ &  0.06 & 13.7\% & 16.9\% &  1.1\arcsec \\
AzTEC\_J021825.18-050923.2 & AzTEC/SXDF25 &  4.1 & $ 7.1 \pm 1.8$ & $ 4.2 \pm ^{+ 1.9}_{- 2.2}$ &  0.03 & 22.5\% & 17.1\% &  0.7\arcsec \\
AzTEC\_J021832.28-045631.3 & AzTEC/SXDF26 &  3.8 & $ 4.7 \pm 1.2$ & $ 3.5 \pm ^{+ 1.9}_{- 2.2}$ &  0.05 & 15.2\% & 16.9\% &  2.6\arcsec \\
AzTEC\_J021838.80-043452.5 & AzTEC/SXDF27 &  3.8 & $ 6.6 \pm 1.7$ & $ 3.6 \pm ^{+ 1.9}_{- 2.1}$ &  0.04 & 17.6\% & 16.9\% &  1.5\arcsec \\
AzTEC\_J021802.43-050019.2 & AzTEC/SXDF28 &  3.8 & $ 4.9 \pm 1.3$ & $ 3.1 \pm ^{+ 1.4}_{- 1.5}$ &  0.02 & 18.6\% & 17.1\% &  1.0\arcsec \\
AzTEC\_J021818.75-045033.4 & AzTEC/SXDF29$^b$ &  4.1 & $ 5.2 \pm 1.3$ & $ 3.5 \pm ^{+ 1.4}_{- 1.4}$ &  0.01 & 31.8\% & 17.0\% & -- \\
AzTEC\_J021826.34-044434.8 & AzTEC/SXDF30$^b$ &  4.0 & $ 5.6 \pm 1.4$ & $ 3.6 \pm ^{+ 1.6}_{- 1.5}$ &  0.01 & 30.6\% & 17.2\% & -- \\
AzTEC\_J021741.41-050217.3 & AzTEC/SXDF31$^b$ &  3.9 & $ 5.4 \pm 1.4$ & $ 3.5 \pm ^{+ 1.5}_{- 1.6}$ &  0.02 & 26.3\% & 17.0\% & -- \\
AzTEC\_J021713.12-045856.8 & AzTEC/SXDF32$^b$ &  3.9 & $ 6.0 \pm 1.5$ & $ 3.6 \pm ^{+ 1.7}_{- 1.8}$ &  0.02 & 27.2\% & 17.2\% & -- \\
AzTEC\_J021737.27-044802.4 & AzTEC/SXDF33$^b$ &  3.9 & $ 5.0 \pm 1.3$ & $ 3.3 \pm ^{+ 1.4}_{- 1.5}$ &  0.02 & 25.7\% & 17.2\% & -- \\
AzTEC\_J021833.86-051019.1 & AzTEC/SXDF34$^b$ &  3.8 & $ 7.3 \pm 1.9$ & $ 3.6 \pm ^{+ 2.2}_{- 2.5}$ &  0.06 & 24.2\% & 17.1\% & -- \\
AzTEC\_J021749.12-045057.5 & AzTEC/SXDF35$^b$ &  3.8 & $ 4.8 \pm 1.3$ & $ 3.1 \pm ^{+ 1.4}_{- 1.4}$ &  0.02 & 24.4\% & 17.1\% & -- \\
AzTEC\_J021656.60-044027.1 & AzTEC/SXDF36$^b$ &  3.8 & $ 6.6 \pm 1.7$ & $ 3.6 \pm ^{+ 1.9}_{- 2.3}$ &  0.05 & 22.7\% & 17.1\% & -- \\
AzTEC\_J021815.94-051255.1 & AzTEC/SXDF37$^b$ &  3.8 & $ 7.3 \pm 1.9$ & $ 3.6 \pm ^{+ 2.0}_{- 2.7}$ &  0.06 & 19.1\% & 17.3\% & -- \\
AzTEC\_J021806.87-044940.5 & AzTEC/SXDF38$^b$ &  3.8 & $ 4.7 \pm 1.2$ & $ 3.0 \pm ^{+ 1.3}_{- 1.5}$ &  0.02 & 19.7\% & 17.1\% & -- \\
AzTEC\_J021809.35-042801.6 & AzTEC/SXDF39$^b$ &  3.8 & $ 6.4 \pm 1.7$ & $ 3.4 \pm ^{+ 1.9}_{- 2.1}$ &  0.05 & 23.1\% & 16.9\% & -- \\
AzTEC\_J021730.94-045133.1 & AzTEC/SXDF40$^b$ &  3.7 & $ 4.7 \pm 1.3$ & $ 2.9 \pm ^{+ 1.4}_{- 1.5}$ &  0.03 & 25.1\% & 17.1\% & -- \\
AzTEC\_J021816.73-050309.5 & AzTEC/SXDF41$^b$ &  3.7 & $ 5.3 \pm 1.4$ & $ 3.0 \pm ^{+ 1.6}_{- 1.7}$ &  0.03 & 33.4\% & 17.0\% & -- \\
AzTEC\_J021740.42-045501.4 & AzTEC/SXDF42$^b$ &  3.7 & $ 4.6 \pm 1.3$ & $ 2.9 \pm ^{+ 1.4}_{- 1.5}$ &  0.03 & 22.8\% & 17.2\% & -- \\
AzTEC\_J021826.64-044933.1 & AzTEC/SXDF43$^b$ &  3.7 & $ 4.7 \pm 1.3$ & $ 2.9 \pm ^{+ 1.4}_{- 1.5}$ &  0.03 & 19.1\% & 17.0\% & -- \\
AzTEC\_J021756.38-045243.1 & AzTEC/SXDF44$^b$ &  3.7 & $ 4.5 \pm 1.2$ & $ 2.8 \pm ^{+ 1.4}_{- 1.4}$ &  0.03 & 14.2\% & 17.0\% & -- \\
AzTEC\_J021858.10-044749.0 & AzTEC/SXDF45$^b$ &  3.7 & $ 5.6 \pm 1.5$ & $ 3.1 \pm ^{+ 1.7}_{- 1.9}$ &  0.04 & 27.2\% & 16.9\% & -- \\
AzTEC\_J021813.11-043810.6 & AzTEC/SXDF46$^b$ &  3.7 & $ 5.8 \pm 1.6$ & $ 3.1 \pm ^{+ 1.7}_{- 2.0}$ &  0.05 & 28.7\% & 17.2\% & -- \\
AzTEC\_J021902.90-045454.3 & AzTEC/SXDF47$^b$ &  3.7 & $ 6.7 \pm 1.8$ & $ 3.2 \pm ^{+ 1.9}_{- 2.5}$ &  0.07 & 22.3\% & 17.3\% & -- \\
AzTEC\_J021730.68-045938.9 & AzTEC/SXDF48$^b$ &  3.7 & $ 4.9 \pm 1.3$ & $ 2.9 \pm ^{+ 1.5}_{- 1.6}$ &  0.03 & 15.7\% & 17.2\% & -- \\
AzTEC\_J021742.60-043859.0 & AzTEC/SXDF49$^b$ &  3.7 & $ 5.5 \pm 1.5$ & $ 3.0 \pm ^{+ 1.7}_{- 1.9}$ &  0.05 & 20.1\% & 17.3\% & -- \\
AzTEC\_J021727.36-050641.0 & AzTEC/SXDF50$^b$ &  3.6 & $ 6.3 \pm 1.7$ & $ 3.1 \pm ^{+ 1.9}_{- 2.3}$ &  0.06 & 22.9\% & 16.9\% & -- \\
AzTEC\_J021833.22-045808.9 & AzTEC/SXDF51$^b$ &  3.6 & $ 4.6 \pm 1.3$ & $ 2.8 \pm ^{+ 1.4}_{- 1.5}$ &  0.03 & 21.5\% & 17.2\% & -- \\
AzTEC\_J021725.32-043845.1 & AzTEC/SXDF52$^b$ &  3.6 & $ 6.1 \pm 1.7$ & $ 3.0 \pm ^{+ 1.8}_{- 2.2}$ &  0.06 & 23.8\% & 17.2\% & -- \\
AzTEC\_J021752.04-050450.8 & AzTEC/SXDF53$^b$ &  3.6 & $ 5.3 \pm 1.5$ & $ 2.9 \pm ^{+ 1.6}_{- 1.8}$ &  0.04 & 20.1\% & 17.0\% & -- \\
AzTEC\_J021808.41-050603.3 & AzTEC/SXDF54$^b$ &  3.6 & $ 5.5 \pm 1.5$ & $ 2.9 \pm ^{+ 1.7}_{- 1.9}$ &  0.05 & 23.2\% & 17.0\% & -- \\
AzTEC\_J021848.92-044204.8 & AzTEC/SXDF55$^b$ &  3.6 & $ 6.0 \pm 1.7$ & $ 2.9 \pm ^{+ 1.9}_{- 2.2}$ &  0.07 & 18.1\% & 17.2\% & -- \\

\hline
\multicolumn{9}{l}{Columns are as described in Table \ref{tbl:cosmos}.}
\end{tabular}
\label{tbl:sxdf}
\end{table*}

\begin{table*}
\contcaption{}
\begin{tabular}{|l|l|c|c|c|c|c|c|c|}
\hline \hline
Source ID & Nickname & S/N & $S_{1.1\text{mm}}$ & $S^\text{corrected}_{1.1\text{mm}}$ & $P(<0)$ & Flux & Noise & $\theta$\\
& & & [mJy] & [mJy] & & increase & increase & \\
\hline
AzTEC\_J021729.76-050325.1 & AzTEC/SXDF56$^b$ &  3.6 & $ 5.3 \pm 1.5$ & $ 2.9 \pm ^{+ 1.6}_{- 1.8}$ &  0.05 & 18.1\% & 17.1\% & -- \\
AzTEC\_J021807.09-043755.1 & AzTEC/SXDF57$^b$ &  3.6 & $ 5.6 \pm 1.6$ & $ 2.9 \pm ^{+ 1.7}_{- 2.0}$ &  0.06 & 32.8\% & 17.1\% & -- \\
AzTEC\_J021759.90-044729.1 & AzTEC/SXDF58$^b$ &  3.6 & $ 4.5 \pm 1.3$ & $ 2.7 \pm ^{+ 1.4}_{- 1.5}$ &  0.04 & 21.3\% & 17.1\% & -- \\
AzTEC\_J021848.94-050015.3 & AzTEC/SXDF59$^b$ &  3.6 & $ 5.9 \pm 1.7$ & $ 2.9 \pm ^{+ 1.8}_{- 2.2}$ &  0.07 & 18.3\% & 17.1\% & -- \\
AzTEC\_J021806.92-044415.3 & AzTEC/SXDF60$^b$ &  3.6 & $ 4.8 \pm 1.3$ & $ 2.7 \pm ^{+ 1.5}_{- 1.6}$ &  0.04 & 12.6\% & 17.1\% & -- \\
AzTEC\_J021752.23-045854.7 & AzTEC/SXDF61$^b$ &  3.6 & $ 4.6 \pm 1.3$ & $ 2.7 \pm ^{+ 1.4}_{- 1.6}$ &  0.04 & 13.9\% & 17.1\% & -- \\
AzTEC\_J021730.04-050951.0 & AzTEC/SXDF62$^b$ &  3.5 & $ 6.6 \pm 1.9$ & $ 2.9 \pm ^{+ 1.2}_{- 3.0}$ &  0.09 & 13.0\% & 17.1\% & -- \\
AzTEC\_J021711.58-045752.8 & AzTEC/SXDF63$^b$ &  3.5 & $ 5.3 \pm 1.5$ & $ 2.7 \pm ^{+ 1.7}_{- 1.9}$ &  0.06 & 14.0\% & 17.1\% & -- \\
AzTEC\_J021743.59-050312.6 & AzTEC/SXDF64$^b$ &  3.5 & $ 4.9 \pm 1.4$ & $ 2.7 \pm ^{+ 1.5}_{- 1.8}$ &  0.05 & 18.8\% & 17.2\% & -- \\
AzTEC\_J021757.59-050035.2 & AzTEC/SXDF65$^b$ &  3.5 & $ 4.5 \pm 1.3$ & $ 2.5 \pm ^{+ 1.5}_{- 1.5}$ &  0.04 & 16.8\% & 17.1\% & -- \\

\hline
\multicolumn{9}{l}{Columns are as described in Table \ref{tbl:cosmos}.}
\end{tabular}
\label{tbl:lh}
\end{table*}

\subsection{GOODS South / ASTE}
The GOODS South field is another commonly observed field and among the first chosen for AzTEC when it was moved from
the JCMT to the ASTE in 2007. A revision of the confusion-limited catalogue presented in \cite{ScottGOODSS2010} is shown in 
Table \ref{tbl:goodss}. Sources of significance greater than 3.5 in the 50 per cent coverage region (0.08 deg$^2$) are presented.
Using the same false detection estimate technique as was used for the COSMOS field, we estimate a false detection rate of
6 per cent ($\sim$3 sources).  Many of the sources appear to be somewhat extended, a sign of the highly confused nature of the
map. Notably, we present the 2nd most significant detection under the assumptions that the flux is the result of a single source or,
alternatively, from two nearby sources. Other sources at modest signal-to-noise do not provide sufficient constraints to multiple
source models. 8 new sources are found in the revised catalogue, many from regions of the map which appear extended.

\begin{table*}
\caption{The AzTEC point source catalogue for the GOODS South field.}
\begin{tabular}{|l|l|c|c|c|c|c|c|c|}
\hline \hline
Source ID & Nickname & S/N & $S_{1.1\text{mm}}$ & $S^\text{corrected}_{1.1\text{mm}}$ & $P(<0)$ & Flux & Noise & $\theta$\\
& & & [mJy] & [mJy] & & increase & increase & \\
\hline
AzTEC\_J033211.48-275216.7 & AzTEC/GS1 & 11.3 & $ 6.9 \pm 0.6$ & $ 6.6 \pm ^{+ 0.6}_{- 0.6}$ &  0.00 &  5.9\% &  8.8\% &  0.7\arcsec \\
AzTEC\_J033218.49-275222.6 & AzTEC/GS2 & 10.6 & $ 6.0 \pm 0.6$ & $ 5.8 \pm ^{+ 0.6}_{- 0.6}$ &  0.00 &  1.2\% &  8.8\% &  0.8\arcsec \\
AzTEC\_J033219.00-275214.6 & \hspace{11.4mm}GS2.1 & 10.2 & $6.8 \pm 0.7$ & $ 6.4 \pm ^{+ 0.7}_{- 0.6}$ &  0.00 & -- & -- & -- \\
AzTEC\_J033216.96-275241.9 & \hspace{11.4mm}GS2.2 & 6.6 & $4.4 \pm 0.7$ & $ 4.0 \pm ^{+ 0.6}_{- 0.7}$ &  0.00 & -- & -- & -- \\
AzTEC\_J033247.70-275419.6 & AzTEC/GS3 &  9.2 & $ 5.1 \pm 0.6$ & $ 4.8 \pm ^{+ 0.6}_{- 0.6}$ &  0.00 &  6.2\% &  8.6\% &  2.1\arcsec \\
AzTEC\_J033248.78-274249.9 & AzTEC/GS4 &  8.7 & $ 5.5 \pm 0.6$ & $ 5.1 \pm ^{+ 0.6}_{- 0.6}$ &  0.00 &  9.5\% &  8.8\% &  0.4\arcsec \\
AzTEC\_J033151.43-274434.5 & AzTEC/GS5 &  7.5 & $ 5.1 \pm 0.7$ & $ 4.7 \pm ^{+ 0.7}_{- 0.7}$ &  0.00 &  5.7\% & 10.6\% &  5.1\arcsec \\
AzTEC\_J033225.73-275219.7 & AzTEC/GS6 &  6.9 & $ 3.8 \pm 0.6$ & $ 3.6 \pm ^{+ 0.6}_{- 0.6}$ &  0.00 & 12.9\% &  8.8\% &  0.3\arcsec \\
AzTEC\_J033213.50-275607.4 & AzTEC/GS7 &  6.5 & $ 4.1 \pm 0.6$ & $ 3.7 \pm ^{+ 0.6}_{- 0.6}$ &  0.00 &  6.6\% &  8.8\% &  0.7\arcsec \\
AzTEC\_J033205.16-274643.8 & AzTEC/GS8 &  6.5 & $ 3.7 \pm 0.6$ & $ 3.4 \pm ^{+ 0.5}_{- 0.6}$ &  0.00 &  6.0\% &  8.5\% &  2.1\arcsec \\
AzTEC\_J033302.60-275149.1 & AzTEC/GS9 &  6.6 & $ 4.0 \pm 0.6$ & $ 3.6 \pm ^{+ 0.6}_{- 0.6}$ &  0.00 &  9.4\% &  8.7\% &  3.1\arcsec \\
AzTEC\_J033207.19-275125.7 & AzTEC/GS10 &  6.3 & $ 4.3 \pm 0.7$ & $ 3.8 \pm ^{+ 0.7}_{- 0.7}$ &  0.00 &  9.2\% &  8.8\% &  0.8\arcsec \\
AzTEC\_J033215.79-275040.2 & AzTEC/GS11 &  6.1 & $ 3.7 \pm 0.6$ & $ 3.3 \pm ^{+ 0.6}_{- 0.6}$ &  0.00 &  6.5\% &  8.7\% &  3.4\arcsec \\
AzTEC\_J033229.33-275616.5 & AzTEC/GS12 &  6.0 & $ 3.5 \pm 0.6$ & $ 3.1 \pm ^{+ 0.6}_{- 0.6}$ &  0.00 &  5.8\% &  9.0\% &  3.8\arcsec \\
AzTEC\_J033211.93-274616.7 & AzTEC/GS13 &  6.2 & $ 3.4 \pm 0.6$ & $ 3.1 \pm ^{+ 0.6}_{- 0.5}$ &  0.00 &  9.7\% &  8.8\% &  0.4\arcsec \\
AzTEC\_J033234.55-275219.5 & AzTEC/GS14 &  6.1 & $ 3.3 \pm 0.5$ & $ 3.0 \pm ^{+ 0.6}_{- 0.6}$ &  0.00 & 10.4\% &  8.7\% &  3.1\arcsec \\
AzTEC\_J033150.93-274601.3 & AzTEC/GS15 &  6.0 & $ 4.4 \pm 0.7$ & $ 3.9 \pm ^{+ 0.7}_{- 0.8}$ &  0.00 &  8.8\% &  8.8\% &  0.9\arcsec \\
AzTEC\_J033237.50-274358.9 & AzTEC/GS16 &  5.6 & $ 3.0 \pm 0.5$ & $ 2.7 \pm ^{+ 0.6}_{- 0.6}$ &  0.00 &  5.9\% &  8.5\% &  3.7\arcsec \\
AzTEC\_J033222.56-274816.5 & AzTEC/GS17 &  5.6 & $ 3.4 \pm 0.6$ & $ 3.0 \pm ^{+ 0.6}_{- 0.6}$ &  0.00 &  8.1\% &  8.8\% &  3.3\arcsec \\
AzTEC\_J033243.60-274634.9 & AzTEC/GS18 &  5.8 & $ 3.5 \pm 0.6$ & $ 3.2 \pm ^{+ 0.6}_{- 0.6}$ &  0.00 & 14.5\% &  8.8\% &  2.1\arcsec \\
AzTEC\_J033223.27-274131.5 & AzTEC/GS19 &  5.4 & $ 3.0 \pm 0.5$ & $ 2.6 \pm ^{+ 0.5}_{- 0.6}$ &  0.00 &  7.4\% &  8.4\% &  2.8\arcsec \\
AzTEC\_J033235.02-275537.7 & AzTEC/GS20 &  5.4 & $ 3.1 \pm 0.6$ & $ 2.7 \pm ^{+ 0.6}_{- 0.6}$ &  0.00 & 12.2\% &  8.7\% &  2.8\arcsec \\
AzTEC\_J033247.45-274443.9 & AzTEC/GS21 &  5.1 & $ 3.1 \pm 0.6$ & $ 2.7 \pm ^{+ 0.6}_{- 0.6}$ &  0.00 &  9.0\% &  8.2\% &  5.8\arcsec \\
AzTEC\_J033212.42-274258.5 & AzTEC/GS22 &  4.6 & $ 2.5 \pm 0.5$ & $ 2.2 \pm ^{+ 0.5}_{- 0.6}$ &  0.00 &  6.9\% &  8.9\% &  2.5\arcsec \\
AzTEC\_J033221.42-275628.4 & AzTEC/GS23 &  4.7 & $ 2.7 \pm 0.6$ & $ 2.3 \pm ^{+ 0.6}_{- 0.6}$ &  0.00 &  9.4\% &  8.8\% &  0.7\arcsec \\
AzTEC\_J033234.55-274949.6 & AzTEC/GS24 &  4.7 & $ 2.8 \pm 0.6$ & $ 2.4 \pm ^{+ 0.6}_{- 0.6}$ &  0.00 & 10.5\% &  8.4\% &  7.1\arcsec \\
AzTEC\_J033246.97-275128.4 & AzTEC/GS25 &  4.3 & $ 2.3 \pm 0.5$ & $ 2.0 \pm ^{+ 0.5}_{- 0.6}$ &  0.00 &  5.3\% &  8.5\% &  6.0\arcsec \\
AzTEC\_J033216.00-274337.6 & AzTEC/GS26 &  4.8 & $ 2.6 \pm 0.5$ & $ 2.2 \pm ^{+ 0.6}_{- 0.5}$ &  0.00 & 20.2\% &  8.8\% &  2.9\arcsec \\
AzTEC\_J033242.43-274155.1 & AzTEC/GS27 &  4.6 & $ 2.6 \pm 0.6$ & $ 2.2 \pm ^{+ 0.6}_{- 0.6}$ &  0.00 & 15.8\% &  8.6\% &  3.3\arcsec \\
AzTEC\_J033242.52-275213.4 & AzTEC/GS28 &  4.6 & $ 2.5 \pm 0.5$ & $ 2.1 \pm ^{+ 0.5}_{- 0.6}$ &  0.00 & 17.6\% &  8.8\% &  7.1\arcsec \\
AzTEC\_J033159.05-274501.1 & AzTEC/GS29 &  4.6 & $ 2.7 \pm 0.6$ & $ 2.3 \pm ^{+ 0.6}_{- 0.6}$ &  0.00 & 20.2\% &  8.3\% &  3.8\arcsec \\
AzTEC\_J033220.78-274240.6 & AzTEC/GS30 &  4.1 & $ 2.2 \pm 0.5$ & $ 1.8 \pm ^{+ 0.6}_{- 0.5}$ &  0.00 &  8.4\% &  8.8\% &  2.1\arcsec \\
AzTEC\_J033242.92-273925.8 & AzTEC/GS31 &  4.2 & $ 2.8 \pm 0.7$ & $ 2.2 \pm ^{+ 0.7}_{- 0.7}$ &  0.00 &  9.9\% &  8.3\% &  1.9\arcsec \\
AzTEC\_J033309.16-275128.3 & AzTEC/GS32 &  4.1 & $ 3.0 \pm 0.7$ & $ 2.4 \pm ^{+ 0.8}_{- 0.8}$ &  0.00 &  7.7\% &  7.8\% &  2.6\arcsec \\
AzTEC\_J033249.45-275316.4 & AzTEC/GS33 &  4.4 & $ 2.3 \pm 0.5$ & $ 2.0 \pm ^{+ 0.5}_{- 0.5}$ &  0.00 & 18.4\% &  9.1\% &  5.6\arcsec \\
AzTEC\_J033229.59-274311.3 & AzTEC/GS34 &  4.0 & $ 2.1 \pm 0.5$ & $ 1.7 \pm ^{+ 0.5}_{- 0.6}$ &  0.00 &  8.9\% &  8.8\% &  3.0\arcsec \\
AzTEC\_J033227.10-274052.5 & AzTEC/GS35 &  4.4 & $ 2.5 \pm 0.6$ & $ 2.1 \pm ^{+ 0.6}_{- 0.6}$ &  0.00 & 20.1\% &  8.6\% &  2.7\arcsec \\
AzTEC\_J033213.97-275516.8 & AzTEC/GS36 &  4.1 & $ 2.5 \pm 0.6$ & $ 2.0 \pm ^{+ 0.6}_{- 0.6}$ &  0.00 & 18.8\% &  8.5\% &  2.9\arcsec \\
AzTEC\_J033256.49-274616.1 & AzTEC/GS37 &  3.8 & $ 2.8 \pm 0.7$ & $ 2.1 \pm ^{+ 0.8}_{- 0.8}$ &  0.00 & 10.6\% &  8.3\% &  5.8\arcsec \\
AzTEC\_J033209.23-274243.8 & AzTEC/GS38 &  3.9 & $ 2.2 \pm 0.6$ & $ 1.8 \pm ^{+ 0.6}_{- 0.6}$ &  0.00 & 19.2\% &  9.0\% &  1.8\arcsec \\
AzTEC\_J033154.32-274537.4 & AzTEC/GS39$^a$ &  3.2 & $ 2.1 \pm 0.7$ & -- & -- & -0.2\% &  8.8\% & -- \\
AzTEC\_J033200.41-274634.6 & AzTEC/GS40 &  3.8 & $ 2.3 \pm 0.6$ & $ 1.8 \pm ^{+ 0.6}_{- 0.6}$ &  0.00 & 18.2\% &  8.7\% &  0.4\arcsec \\
AzTEC\_J033302.26-275648.7 & AzTEC/GS41$^c$ &  8.2 & $ 7.9 \pm 1.0$ & $ 7.2 \pm ^{+ 0.9}_{- 1.0}$ &  0.00 & 10.9\% &  8.1\% &  0.4\arcsec \\
AzTEC\_J033314.32-275608.0 & AzTEC/GS42$^c$ &  7.9 & $10.2 \pm 1.3$ & $ 9.2 \pm ^{+ 1.2}_{- 1.4}$ &  0.00 & 11.4\% &  9.3\% &  0.4\arcsec \\
AzTEC\_J033303.05-274428.6 & AzTEC/GS43$^c$ &  6.7 & $ 6.9 \pm 1.0$ & $ 6.1 \pm ^{+ 1.1}_{- 1.1}$ &  0.00 &  3.6\% &  8.0\% &  0.4\arcsec \\
AzTEC\_J033240.81-273801.5 & AzTEC/GS44$^c$ &  4.9 & $ 3.9 \pm 0.8$ & $ 3.3 \pm ^{+ 0.8}_{- 0.8}$ &  0.00 &  5.8\% &  8.7\% &  0.4\arcsec \\
AzTEC\_J033219.15-273733.2 & AzTEC/GS45$^c$ &  4.7 & $ 5.2 \pm 1.1$ & $ 4.2 \pm ^{+ 1.1}_{- 1.3}$ &  0.00 &  9.9\% & 10.5\% &  0.4\arcsec \\
AzTEC\_J033157.27-275702.4 & AzTEC/GS46$^c$ &  4.6 & $ 6.4 \pm 1.4$ & $ 4.8 \pm ^{+ 1.5}_{- 1.6}$ &  0.00 & 10.1\% & 12.7\% &  0.4\arcsec \\
AzTEC\_J033208.21-275821.7 & AzTEC/GS47$^c$ &  4.8 & $ 4.3 \pm 0.9$ & $ 3.6 \pm ^{+ 0.9}_{- 1.0}$ &  0.00 & 13.8\% &  9.0\% &  0.4\arcsec \\
AzTEC\_J033215.81-275249.7 & AzTEC/GS48$^b$ &  4.3 & $ 2.5 \pm 0.6$ & $ 2.1 \pm ^{+ 0.6}_{- 0.6}$ &  0.00 &  5.0\% &  8.8\% & -- \\
AzTEC\_J033221.67-274013.4 & AzTEC/GS49$^b$ &  4.1 & $ 2.6 \pm 0.6$ & $ 2.1 \pm ^{+ 0.6}_{- 0.7}$ &  0.00 & 27.6\% &  8.8\% & -- \\
AzTEC\_J033235.02-274926.0 & AzTEC/GS50$^b$ &  4.1 & $ 2.4 \pm 0.6$ & $ 2.0 \pm ^{+ 0.6}_{- 0.6}$ &  0.00 &  4.3\% &  8.8\% & -- \\
AzTEC\_J033157.52-274507.6 & AzTEC/GS51$^b$ &  3.7 & $ 2.2 \pm 0.6$ & $ 1.7 \pm ^{+ 0.6}_{- 0.6}$ &  0.01 & 10.0\% &  8.8\% & -- \\
AzTEC\_J033244.08-275013.7 & AzTEC/GS52$^b$ &  3.6 & $ 2.0 \pm 0.6$ & $ 1.6 \pm ^{+ 0.6}_{- 0.6}$ &  0.01 & 24.8\% &  8.8\% & -- \\
AzTEC\_J033204.48-274455.6 & AzTEC/GS53$^b$ &  3.6 & $ 1.9 \pm 0.5$ & $ 1.5 \pm ^{+ 0.6}_{- 0.6}$ &  0.01 & 34.8\% &  8.8\% & -- \\
AzTEC\_J033243.17-275516.5 & AzTEC/GS54$^b$ &  3.5 & $ 2.0 \pm 0.6$ & $ 1.5 \pm ^{+ 0.6}_{- 0.6}$ &  0.01 & 15.6\% &  8.8\% & -- \\
AzTEC\_J033225.70-275829.1 & AzTEC/GS55$^b$ &  3.5 & $ 2.6 \pm 0.8$ & $ 1.9 \pm ^{+ 0.8}_{- 0.8}$ &  0.01 & 20.1\% &  8.8\% & -- \\

\hline
\multicolumn{9}{l}{Columns are as described in Table \ref{tbl:cosmos}. ($^c$) indicates a source found in the extended, lower coverage regions of the map.}
\label{tbl:goodss}
\end{tabular}
\end{table*}
\subsection{Observations}
The detection significance of any given source may change owing to the greater accuracy of flux and noise estimation using the 
revised photometry technique. Nonetheless, viewing the catalogues as a whole, detection significance is seen not to be greatly 
impacted because the photometry affects both signal and noise. Likewise, the false detection rates in the fields were unchanged.
The vast majority of sources that passed a significance test in a prior catalogue pass it again in the revised catalogue. In fact, the
increased number of sources in the majority of the revised catalogues suggest a slight systematic upward shift in detection
significance; the exponentially increasing number of dim sources in these fields allow a slight shift to increase considerably the
number of detected sources.

\section{Future Work}
The reviewer suggested a potentially simpler technique for estimating the transfer function that would not require the noise
mitigating steps described. Rather than calculating the transfer function by differencing a source-added map from the
recorded map, one would calculate the eigenvectors from the source-added timestreams and remove them from source-only
timestreams. The eigenfunctions identified by PCA cleaning would have the proper impact of sources included, but the transfer
function estimated would not include the small difference in noise realisations between the two maps. This should produce a
result identical to the analysis shown in a much simpler fashion. This manuscript documents the technique used in
several AzTEC publications in 2011-2012 \citep{Humphrey2011, Yun2011, Aretxaga2011, Kim2012}. If the technique is revised further
to include this suggestion or others, it will be documented in subsequent publications.  

\section{Conclusions}
A general approach to estimating the transfer function of non-linear techniques has been described and applied to the 
specific case of principal component analysis (PCA). Simulations support the accuracy of the results and that 
PCA has a transfer function which is effectively linear for point sources of typical detection significance. The resulting transfer
function has been used to correct the catalogue values for the flux, location and significance of point sources in existing AzTEC
maps. Mean source detection significance is not strongly impacted by the photometry correction and may be slightly enhanced.

\section{Acknowledgments}
This work has been made possible by generous support from the Kavli Foundation and the Gordon and Betty Moore Foundation. 
Additional support for this analysis has been provided in part by National Science Foundation grants 0540852, 0838222 and 
0907952. KSS is supported by the National Radio Astronomy Observatory, which is a facility of the National Science Foundation
operated under cooperative agreement by Associated Universities, Inc. We would like to acknowledge the assistance of Jack Sayers
and Stephan Meyer for useful comments on the analysis presented herein. We also thank the observatory staff of the JCMT and
ASTE who made these observations possible.

\bibliographystyle{mn2e}
\bibliography{references}

\begin{thebibliography}{}

\bibitem[\protect\citeauthoryear{Anton}{Anton}{1994}]{linalg}
Anton H.,  1994, Elementary Linear Algebra, 7th edn.
Wiley, New York

\bibitem[\protect\citeauthoryear{{Aretxaga} et~al.,}{{Aretxaga}
  et~al.}{2011}]{Aretxaga2011}
{Aretxaga} I.,  et~al., 2011, \mnras, 415, 3831

\bibitem[\protect\citeauthoryear{{Austermann} et~al.,}{{Austermann}
  et~al.}{2009}]{AustermannCOSMOS2009}
{Austermann} J.~E.,  et~al., 2009, \mnras, 393, 1573

\bibitem[\protect\citeauthoryear{{Austermann} et~al.,}{{Austermann}
  et~al.}{2010}]{AustermannSHADES2010}
{Austermann} J.~E.,  et~al., 2010, \mnras, 401, 160

\bibitem[\protect\citeauthoryear{Enoch et~al.,}{Enoch
  et~al.}{2006}]{Enoch2006}
Enoch M.~L.,  et~al., 2006, \apj, 638, 293

\bibitem[\protect\citeauthoryear{{Humphrey} et~al.,}{{Humphrey}
  et~al.}{2011}]{Humphrey2011}
{Humphrey} A.,  et~al., 2011, \mnras, 418, 74

\bibitem[\protect\citeauthoryear{{Kim} et~al.,}{{Kim}  et~al.}{2012}]{Kim2012}
{Kim} M.~J.,  et~al., 2012, \apj, 746, 11

\bibitem[\protect\citeauthoryear{{Kov{\'a}cs}}{{Kov{\'a}cs}}{2008}]{KovacsCRUS%
H2008}
{Kov{\'a}cs} A.,  2008, in Society of Photo-Optical Instrumentation Engineers
  (SPIE) Conference Series Vol.~7020 of Presented at the Society of
  Photo-Optical Instrumentation Engineers (SPIE) Conference, {CRUSH: fast and
  scalable data reduction for imaging arrays}

\bibitem[\protect\citeauthoryear{{Lay} \& {Halverson}}{{Lay} \&
  {Halverson}}{2000}]{LayHalverson2000}
{Lay} O.~P.,  {Halverson} N.~W.,  2000, \apj, 543, 787

\bibitem[\protect\citeauthoryear{{Perera} et~al.,}{{Perera}
  et~al.}{2008}]{PereraGOODSN2008}
{Perera} T.~A.,  et~al., 2008, \mnras, 391, 1227

\bibitem[\protect\citeauthoryear{Sayers}{Sayers}{2007}]{SayersThesis}
Sayers J.,  2007, PhD thesis, California Institute of Technology

\bibitem[\protect\citeauthoryear{{Sayers} et~al.,}{{Sayers}
  et~al.}{2010}]{SayersMK2010}
{Sayers} J.,  et~al., 2010, \apj, 708, 1674

\bibitem[\protect\citeauthoryear{{Scott} et~al.,}{{Scott}
  et~al.}{2008}]{ScottCOSMOS2008}
{Scott} K.~S.,  et~al., 2008, \mnras, 385, 2225

\bibitem[\protect\citeauthoryear{{Scott} et~al.,}{{Scott}
  et~al.}{2010}]{ScottGOODSS2010}
{Scott} K.~S.,  et~al., 2010, \mnras, 405, 2260

\bibitem[\protect\citeauthoryear{{Wilson} et~al.,}{{Wilson}
  et~al.}{2008}]{WilsonAztecInstrument}
{Wilson} G.~W.,  et~al., 2008, \mnras, 386, 807

\bibitem[\protect\citeauthoryear{{Yun} et~al.,}{{Yun}  et~al.}{2011}]{Yun2011}
{Yun} M.~S.,  et~al., 2011, \mnras, p.~2141

\end{thebibliography}

\end{document}